\documentclass[12pt]{article}
\usepackage[pdftex]{graphicx}
\usepackage{hyperref}
\usepackage{amsmath}
\usepackage{amssymb}
\renewcommand{\title}[1]{%
    \bigskip%
    \begin{center}%
    \Large\bf #1%
    \end{center}%
    \vskip .2in}

\renewcommand{\author}[1]{%
    {\begin{center}
    #1
    \end{center}}}
\newcommand{\address}[1]{\vspace{-1.7em}\vspace{0pt}
    {\begin{center}
    \it #1
    \end{center}}}

\begin{document}

\title{\bf{Canonical Formulation  for a Non-relativistic Spinning Particle Coupled to Gravity}}

\author
{
Rabin Banerjee  $\,^{\rm a,b}$,
Pradip Mukherjee  $\,^{\rm c,d}$}
\address{$^{\rm a}$ S. N. Bose National Centre 
for Basic Sciences, JD Block, Sector III, Salt Lake City, Kolkata -700 098, India }

\address{$^{\rm c}$Department of Physics, Barasat Government College,Barasat, India}
\address{$^{\rm b}$\tt rabin@bose.res.in}
\address{$^{\rm d}$\tt mukhpradip@gmail.com}

\vskip 1cm
\begin{abstract}
We systematically derive  an action for a nonrelativistic spinning partile in flat background and discuss its canonical formulation in both Lagrangian and Hamiltonian approaches. This action is taken as the starting point for deriving the corresponding action in a curved background. It is achieved by following our recently developed technique of localising the flat space galilean symmetry \cite{BMM1, BMM3, BMM2}. The coupling of the spinning particle to a Newton-Cartan background is obtained naturally. The equation of motion is found to differ from the geodesic equation, in agreement with earlier findings. Results for both the flat space limit and the spinless theory (in curved background) are  reproduced. Specifically, the geodesic equation is also obtained in the latter case. 
\end{abstract}

\section{Introduction }

\smallskip

The study of nonrelativistic (NR) symmetries in a gravitational background has recently ~attracted considerable attention since it has found applications in various topics related to particle physics, condensed matter physics \cite{SW}, \cite{Grin}, \cite{Gr} fluid dynamics \cite {A}, \cite{Mi}, gravitation \cite{M}, \cite{Rea} and cosmology \cite{K}, to name a few. The possibility of such symmetries was already noticed by Cartan \cite{Cartan-1923},\cite {Cartan-1924}
, who developed a covariant geometrical theory of Newtonian gravitation some time after Einstein formulated his general theory of relativity. The corresponding Newton Cartan (NC) manifold has a degenerate metric structure and there exists an extensive literature on this subject \cite{Havas} - \cite{EHL}. In recent applications, the elements of NC geometry are used to couple the matter sector , be it particles \cite{Kuch, PPP, BMN1}, extended objects or fields \cite{BGMM, j1}, with NR gravity. 

The universal role of time in NR physics imposes the lack of a single nondegenerate metric which naturally occurs in the relativistic case. It is then necessary to work with a couple of degenerate structures, which are the elements of NC geometry. Consequently, contrary to the relativistic case, there is no obvious way to couple the matter sector with NC geometry . Among the various approaches discussed in the recent times, some popular ones are based on gauging the NR (Bargmann) algebra \cite{ABPR} or taking the NR limit of relativistic theories \cite{BGMM, j1}. Apart from the fact that the NR reduction of relativistic theories is not  unique as it can be done in different ways \cite{BGMM}, there are other problems, particularly related to achieving the correct flat limit \cite{BMN1, BM10}. 

Recently we have developed a systematic algorithm of coupling NR matter to gravity that is free of such obscurities and has wide applicability \cite{BMM1}, \cite{BMM3}. It is based on localising the NR (galilean) symmetry and naturally leads to a geometric interpretation involving the NC structures \cite{BMM2}. Christened as galilean gauge theory \cite{BM4}, it has been successfully applied in different contexts, reproducing familiar results and also yielding new findings and insights \cite{BM4},  \cite{BM5} - \cite{BMN2}.
In its barest essentials, it answers the question that, given a theory in flat background, what would be the corresponding theory in a curved background.  The original galilean symmetry is localised by making the parameters of the transformation space time dependent. The ordinary derivatives are replaced by suitable covariant derivatives so that the new theory is invariant under the localised transformations. A geometrical interpretation of this new theory is possible by identifying appropriate combinations of the new fields appearing in the covariant derivatives with the elements of NC geometry. In this way NR diffeomorphism invariance appears naturally while the flat limit is smoothly recovered by simply setting the new fields to zero. 

Here we exploit this 
 approach based on galilean gauge theory to discuss the canonical formulation of a nonrelativisitc Fermi (spin half) particle coupled to a curved background.  Conceptually we follow similar steps as done by us for the spinless theory \cite{BMN1, BMN2} but nontrivial technicalities appear due to the inclusion of spin. Before summarising the contents of this paper, we mention that the action principle for relativistic spinning objects and their equations of motion have been considered earlier at great length. Initiated by Mathisson-Papapetrou-Dixon \cite{math, papa, dixon}, this investigation was  generalised by Anandan et al \cite{anandan, anandan1}. A different analysis of a similar problem was  carried out by Hanson, Regge and Teitelboim using 	constrained analysis of a spinning top \cite{HRT}. These studies are, however, all confined to relativistic systems. The present paper addresses the issue of a  nonrelativistic spin half particle. We shall make a comparison of our results with the relativistic ones in due course.

 The first step in our analysis would be to construct the appropriate theory in a flat space time. This has been presented in some details in section 2. To illustrate the method we first consider the simpler and well known example of the free spinless particle. The usual action is reproduced in a parametrisation invariant form by introducing the Schroedinger mass shell condition as a constraint. The action for the free spinning theory in flat space time is then constructed in 
 section 2.1. Now, apart from the Schroedinger mass shell constraint, there is also the Pauli Schroedinger constraint. The canonical formulation is developed in 2.1.1. The equation of motion for the coordinates shows a deviation from the expected straight line path for a free particle. It is, however, possible to reproduce the standard geodesic equation in terms of the momenta. This happens because the momenta are not aligned with the coordinates. The galilean symmetry is examined both for the ordinary and spin (Grassmann) variables in  2.1.2. The action is shown to be quasi-invariant. A Hamiltonian formulation of the action reveals the presence of reparametrisation symmetry and super gauge symmetry. The appropriate generator that yields these symmetries is constructed in section 2.1.3. Also, the method  is extended to include interactions with an external magnetic field in 2.1.4. We show explicitly that the results follow by a minimal substitution, replacing the momenta by including the vector potential term. The flat space action obtained here is the starting point to derive the action in a curved background. This is presented in section 3, using the algorithm of galilean gauge theory. A geometric interpretation of the action is provided in section 3.1. The action is written in a manifestly covariant form where the coupling with NC geometry becomes explicit. In section 4, a Lagrangian analysis is done. It is found that the equation of motion for the coordinates does not follow the geodesic equation, in agreement with recent findings \cite{BCG}. But we are able to go further and provide new insights. This is done in section 5 where a comparison with existing results is performed. For the specific case of a spinning particle in a flat background we are able to obtain the precise nonrelativistic analogues of Papapetrou's equations written in a flat background \cite{papa}. We also comment on possible interpretations for the curved background results.  
 Finally, concluding remarks are given in section 6 where we briefly mention about possible phenomenological consequences.

\section{Free Spinless Particle}

 The dispersion relation connecting the three momentum $p_i$ with the energy 
$E$ is given by,

\begin{equation}
E = \frac{p_i^2}{2m}
\label{dispersion}
\end{equation}

Introducing this as a constraint, the first order form of the action becomes,
\begin{equation}
S = \int d\tau \Big(p_i \dot x^i - E \dot t  +\frac{e}{2} (E -\frac{p_i^2}{2m})\Big)
\label{action}
\end{equation}
where $e$ is a Lagrange multiplier and an overdot indicates a differentiation wuth respect to the arbitrary parameter $\tau$. The symplectic structure is defined by the Poisson brackets,
\begin{equation}
\{x_i, p^j\} = \delta_i^j, \,\,\, \{t, E\} = -1
\label{poisson}
\end{equation}

By using the equations of motion for the variables $p_i$ and $E$, it is possible to eliminate them and simplify the form of the action to,

\begin{equation}
S= \frac{m}{2} \int d\tau \,\,\frac{\dot x_i^2}{\dot t}
\label{action1}
\end{equation}

This is the action for a free spinless non-relativistic particle written in a reparametrisation invariant form. The familiar form is obtained by choosing the standard gauge $t=\tau$,

\begin{equation}
S= \frac{m}{2} \int dt \,\,\dot x_i^2
\label{action111}
\end{equation}
where the overdot here denotes a differentiation with respect to the usual time $t$.{\footnote{Notation: Whether the overdot refers to a differentiation with regard to the the usual time $t$ or the parametrised time $\tau$ can be easily understood from the integration variable. For unintegrated expressions the definition of the overdot will be stated explicitly.}}

In the following section we will introduce spin degrees of freedom. Meanwhile let us note  the invariance of the  action (\ref{action1}) under infinitesmal Galilean transformations,
\begin{eqnarray}
\delta x^0 &=& \theta^0 = -\theta\nonumber\\
\delta x^i &=& \theta^i - \omega^{i}\,_j x^j - v^i t
\label{galtrans}
\end{eqnarray}
where, $\theta^\mu$ are the time and space translation parameters, $\omega_{i j}$ are spatial rotation parameters and $v^i$ are the Galilean boost parameters, 
The corresponding change in the action is,
\begin{equation}
\delta S = \int d\tau \frac{d}{d\tau}\left( - m v_i x_i\right)
\label{quasi1}
\end{equation}
Note that the action varies by a total boundary, keeping the equations of motion invariant. To distinguish this type of invariance from those where the action does not change $(\delta S = 0)$, it is called a 
 quasi invariance.

\subsection{The Spinning Theory}

The action for the spinning particle is constructed in a manner similar to the spinless case. Here, apart from the dispersion relation (\ref{dispersion}), there is another constraint that implies the spinning nature of the particle. Since we are considering the spin half case, this condition is given by the Pauli-Schroedinger equation. One possible way to obtain this is to start from the four component Dirac equation. Expressing the four component Dirac fermion in terms of a pair of two component fermions, taking the nonrelativistic $(c\rightarrow \infty)$ limit after eliminating the oscillations due to the rest mass of the fermions, eventually yields the following set of coupled equations \cite{LL, BC},
\begin{eqnarray}
E\,\psi - (\sigma.p)\,\eta &=& 0 \cr
(\sigma.p)\, \psi - 2m \eta &=& 0
\label{ps}
\end{eqnarray}
where $\eta$ and $\xi$ are two component spinors and $\sigma_i$ are the Pauli matrices. Using the identity,
\begin{equation}
(\sigma.p)\,(\sigma.p) = p^2\label{identity}
\end{equation}
it is simple to see that both spinors satisfy the Schroedinger equation corresponding to the dispersion relation (\ref{dispersion}),
\begin{equation}
(p^2 - 2m E)\,\eta = (p^2 - 2m E)\,\psi = 0\label{dispersion1}
\end{equation}

The pair of coupled equations (\ref{ps}) may be expressed in four component notation by,
\begin{equation}
\Big[
\begin{pmatrix}
0 & 0 \\ E & 0
\end{pmatrix}
+
\begin{pmatrix}
\sigma.p & 0 \\ 0 & -(\sigma.p)
\end{pmatrix}
-
\begin{pmatrix}
0 & 2m \\ 0 & 0
\end{pmatrix}
\Big]
\begin{pmatrix}
\psi & \\ \eta
\end{pmatrix}
= 0
\label{matrix}
\end{equation}
where each individual entry is a 2 by 2 matrix. These equations describe a non-relativistic spin half particle. Further, if we define the 4 by 4 matrix in (\ref{matrix}) by $Q$,
\begin{equation}
Q =\Big[
\begin{pmatrix}
0 & 0 \\ E & 0
\end{pmatrix}
+
\begin{pmatrix}
\sigma.p & 0 \\ 0 & -(\sigma.p)
\end{pmatrix}
-
\begin{pmatrix}
0 & 2m \\ 0 & 0
\end{pmatrix}
\Big]
\label{brs}
\end{equation}
it is found that its anticommutator is just the Schroedinger operator,
\begin{equation}
\frac{1}{2}\{ Q, Q\}_+ = (p^2 - 2 m E)I
\label{brs1}
\end{equation} 
where $I$ is the 4 by 4 unit matrix. 

The above analysis shows that the analogue of the Schroedinger operator, which is the Pauli-Schroedinger (or the Levy -Leblond) operator has to be  Grassmann in nature. It is defined as the operator version of $Q$ given in (\ref{brs}),
\begin{equation}
P= E\eta + 2 p_i \epsilon_i - 2m \bar\eta
\label{LL}
\end{equation}
where $\eta$, $\bar\eta$ and $\epsilon_i$ are Grassmann variables. A particular 4 by 4 matrix representation of these operators is obtained by comparison with (\ref{brs}),
\begin{equation}
\eta =
\begin{pmatrix}
0 & 0 \\
1 & 0
\end{pmatrix}
\,\,
\bar\eta =
\begin{pmatrix}
0 & 1\\
0 & 0
\end{pmatrix}
\,\,
2\epsilon_i = 
\begin{pmatrix}
\sigma_i & 0\\
0 & -\sigma_i
\end{pmatrix}
\label{rep}
\end{equation}
which reproduces the expected Grassmann algebra,
\begin{equation}
\{\eta, \eta\}_+ = \{\bar\eta, \bar\eta\}_+ = 0, \,\, \{\eta, \bar\eta\}_+ =I, \,\, \{\epsilon_i, \epsilon_j\}_+ = \frac{1}{2} \delta_{ij}I
\label{algebra}
\end{equation}
while the remaining algebra involving the $\eta$ and $\bar\eta$ with $\epsilon_i$ also vanishes. Naturally the Pauli Schroedinger operator (\ref{LL})  satisfies an operator analogue of the  relation  (\ref{brs1}),
\begin{equation}
\frac{1}{2}\{ P, P\}_+ = (p^2 - 2 m E)
\label{brs2}
\end{equation}  

Classically, the Grassmann variables satisfy graded Poisson brackets that are obtained from (\ref{algebra}),
\begin{equation}
\{\eta, \eta\} = \{\bar\eta, \bar\eta\}= 0, \,\, \{\eta, \bar\eta\}=-i, \,\, \{\epsilon_i, \epsilon_j\} = -\frac{i}{2} \delta_{ij}
\label{algebra1}
\end{equation}
while the remaining brackets involving cross terms vanish.

We are now ready to write down the first order form of the action for a free non-relativistic particle of spin half, which is given by,
\begin{equation}
S = \int d\tau \Big(p_i \dot x^i - E \dot t  +\frac{e}{2} (E -\frac{p_i^2}{2m}) - \chi ( E\eta + 2 p_i \epsilon_i - 2m \bar\eta) + \frac{i}{2} (\eta \dot{\bar\eta} + \bar\eta \dot\eta) + i\epsilon_i \dot\epsilon_i\Big)
\label{g11n}
\end{equation}
The difference from the spinless action (\ref{action}) is seen from the fourth term onwards. There is a term that enforces the Pauli-Schroedinger operator as a constraint through the multiplier $\chi$. The remaining kinetic terms are such that the symplectic structure (\ref{algebra1}) is reproduced. If we drop the terms involving Grassmann variables which are responsible for the spin structure, we just get back the spinless form (\ref{action}). 

This action will now be expressed in a second order form. This is done to facilitate the passage to a curved background using our approach based on galilean gauge theory. It might be recalled that the starting point of our \cite{BMN1} derivation for the action of a non-relativisitc spinless particle in a curved background was (\ref{action1}).

Variation of the multiplier $e$ simply yields (\ref{dispersion}), which is used to eliminate $E$ from (\ref{g1}),
\begin{equation}
S = \int d\tau \Big(p_i \dot x^i - \frac{p_i^2}{2m} \dot t   - \chi ( \frac{p_i^2}{2m}\eta + 2 p_i \epsilon_i - 2m \bar\eta) + \frac{i}{2} (\eta \dot{\bar\eta} + \bar\eta \dot\eta) + i\epsilon_i \dot\epsilon_i\Big)
\label{g1}
\end{equation}
Variation of $p_i$ now yields,
\begin{equation}
p_i = \frac{m (\dot x_i - 2 \chi \epsilon_i)}{ \dot t + \chi \eta}
\label{p}
\end{equation}
which is used to eliminate it from (\ref{g1}). This leads to,
\begin{equation}
S = \int d\tau \Big(\frac{m}{2} \frac{(\dot x_i - 2 \chi \epsilon_i)^2}{\dot t + \chi\eta} + 2m \chi \bar\eta    + \frac{i}{2} (\eta \dot{\bar\eta} + \bar\eta \dot\eta) + i\epsilon_i \dot\epsilon_i\Big)
\label{spinaction2}
\end{equation}

This is our cherished form of the action. A similar action was earlier posited in \cite{G1}. However, we have given a systematic derivation. As subsequent analysis will show, there are some fine differences. The above action may be simplified further by exploiting the Grassmann nature of the $\chi$-variable $(\chi^2=0)$, so that,
\begin{equation}
S = \int d\tau \Big(\frac{m}{2}\left( \frac{\dot x_i^2 - 4\chi \epsilon_i\dot x_i}{\dot t + \chi\eta}\right) + 2m \chi \bar\eta    + \frac{i}{2} (\eta \dot{\bar\eta} + \bar\eta \dot\eta) + i\epsilon_i \dot\epsilon_i\Big)
\label{spinaction3}
\end{equation}

\subsubsection{Canonical Formulation}

A canonical formulation of the model is now presented, following Dirac's analysis of constrained systems \cite{D}
The canonical momenta corresponding to the coordinates $x_i$, time $t$ and the multiplier $\chi$ are given by,
\begin{eqnarray}
p_i = \frac{\partial L}{\partial \dot x_i} &=& \frac{m(\dot x_i - 2 \chi \epsilon_i)}{\dot t + \chi\eta}\cr
p_0 = -E = \frac{\partial L}{\partial \dot t} &=& \frac{-m(\dot x_i - 2 \chi \epsilon_i)^2}{2(\dot t + \chi\eta)^2}\cr
\Pi_\chi = \frac{\partial L}{\partial \dot \chi} &=& 0
\label{momenta}
\end{eqnarray}

Terms involving $\eta$, $\bar\eta$ and $\epsilon_i$ are first order and naturally lead to the algebra (\ref{algebra1}), hence their corresponding momenta are not introduced.

The Lagrange equations of motion that follow from (\ref{spinaction2}) are now written down,
\begin{eqnarray}
\dot p_i &=& 0\cr
\dot p_0 &=& 0\cr
\dot \epsilon_i &=& i \chi p_i\cr
\dot \eta &=& - 2i m \chi\cr
\dot{\bar\eta} &=& i E \chi
\label{lagrangeequation}
\end{eqnarray}
where the momenta have been already defined in (\ref{momenta}). There are certain relations among the momenta since these are not all independent. Together with the last relation in (\ref{momenta}), these constitute the primary constraints of the theory and are given by,
\begin{eqnarray}
\Phi_1 &=& p_i^2 - 2E m \approx 0\cr
\Phi_2 &=& \Pi_\chi \approx 0
\label{primary}
\end{eqnarray}
The constraints are implemented weakly, following the nomenclature of Dirac \cite{D, HRT},  meaning that they can be imposed only after all relevant brackets have been computed.

The canonical Hamiltonian is given by,
\begin{equation}
H_c = \chi ( E\eta + 2 p_i \epsilon_i - 2 m \bar\eta)
\label{canonical}
\end{equation}
The total Hamiltonian is then defined by adding the primary constraints with appropriate multipliers \cite{D},
\begin{equation}
H_t = H_c + \lambda_1 \Phi_1 + \lambda_2 \Phi_2
\label{total}
\end{equation}

To obtain secondary constraints, if any, it is necessary to check the time conservation of the primary constraints by (graded) Poisson bracketing with the total Hamiltonian.  For $\Phi_1$ this is trivial,
\begin{equation}
\{\Phi_1, H_t \} = 0
\end{equation}
However, for $\Phi_2$, the algebra is nontrivial,
\begin{equation}
\{\Phi_2 , H_t \} = - \Phi_3
\label{consistency}
\end{equation}
where $\Phi_3$ is a new (secondary) constraint,
\begin{equation}
\Phi_3 = E\eta + 2 p_i \epsilon_i - 2 m \bar\eta \approx 0
\label{secondary}
\end{equation}
To check the presence of other constraints, the process has to be repeated for $\Phi_3$. We find,
\begin{equation}
\{\Phi_3 , H_t\} = -2 i \chi \Phi_1
\label{calgebra}
\end{equation}
No new constraint is generated and the iterative process thus terminates. Note that $\Phi_1$ and $\Phi_3$ are just  the Schoedinger and Pauli-Schroedinger operators, now implemented as  constraints. 

Since the algebra of constraints closes,
\begin{equation}
\{\Phi_a, \Phi_b \} = f_{ab}\,^c \Phi_c , \, f_{33}\,^1 = - 2i \,\, (a, b, c =1, 2, 3)
\label{constraint algebra}
\end{equation}
where all other structure constants are vanishing, the constraints are all first class.
 The total Hamiltonian is then  expressed as a sum of the first class constraints only,
 \begin{equation}
H_t = \lambda_1 \Phi_1 + \lambda_2 \Phi_2 +\chi \Phi_3
\label{total1}
 \end{equation}
 as one would expect for a reparametrisation invariant theory \cite{D, HRT}.
 
 The above Hamiltonian reproduces the momenta (\ref{momenta}) as well as the  Lagrange equation of motion 
(\ref{lagrangeequation}) provided we choose the multipliers as,
\begin{eqnarray}
\lambda_1 &=& \frac{1}{2m} ( \dot t + \chi\eta) \cr
\lambda_2 &=& \dot\chi
\label{mult}
\end{eqnarray}
and use the Grassmann condition $\chi^2 =0$. For example, as a consistency,
\begin{equation}
\dot t = \{t, H_t \} = - \chi\eta + 2m \lambda_1 = \dot t
\label{t}
\end{equation}
and likewise for the other variables.

It is instructive to compute the equation of motion for the coordinates since it gives us an idea of the geodesics. This equation is obtained by varying the  action or simply by using the results of (\ref{momenta}) and (\ref{lagrangeequation}) and  is given by,
\begin{equation}
\frac{d}{d\tau}\Big(\frac{\dot x_i - 2 \chi\epsilon_i}{\dot t + \chi\eta}\Big) = 0
\label{newequation}
\end{equation}
To retrieve the equation in the coordinate time we put $t=\tau$ so that,
\begin{equation}
\frac{d}{dt}\Big(\frac{\dot x_i - 2 \chi\epsilon_i}{ 1+ \chi\eta}\Big) = 0
\label{newequation1}
\end{equation}
where the $dot$ now refers to differentiation with respect to the time $t$.

The above equation may be simplified by again exploiting (\ref{lagrangeequation}). It is seen that the derivatives of $\epsilon_i$ and $\eta$ are proportional to $\chi$. Thus terms like $\chi\dot\eta$ and $\chi\dot\epsilon_i$ in (\ref{newequation1}) drop out since they vanish by the Grassmann condition $\chi^2=0$. Using these results the final form for the equation of motion is,
\begin{equation}
\ddot x_i = (1+\chi\eta)^{-1} \dot\chi(\eta \dot x_i + 2\epsilon_i)
\label{flatgeodesic}
\end{equation}

The same result, expectedly, follows from the total hamiltonian (\ref{total1}). Taking the graded Poisson algebra of the coordinates with the total hamiltonian, where the multipliers are fixed from (\ref{mult}), we obtain,
\begin{equation}
\dot x_i = \{x_i, H_t\}= \frac{1}{m}(1+\chi\eta)p_i  + 2\chi\epsilon_i
\end{equation}
where we have put $t=\tau$. Taking a further time derivative and using the form of the momenta given in (\ref{momenta}) as well as $\dot p_i=0$, which follows either from (\ref{lagrangeequation}) or from the Hamiltonian equation of motion, we are immediately led to (\ref{flatgeodesic}). 

We reserve further discussion on the implications of the equations of motion (see section 5) till the studies of spinning particle motion 
 in a curved background.

\subsubsection{Galilean Symmetry}

We now present a detailed study of the galilean symmetry of the model. The galilean transformations for the coordinates $x_i$ , their  momenta $p_i$ and time $t$ along with energy $E$ are known. The corresponding transformations for the Grassmann variables are found by demanding algebraic consistency of relations containing them together with the usual variables. We show this by an explicit example. Consider the rotation generator for the usual sector,
\begin{equation}
R = \frac{1}{2} \omega_{ij} (x_i p_j  - x_j p_i)
\label{rotation}
\end{equation} 
Now consider its action on the constraint (\ref{secondary}). Under infinitesimal  rotations,
\begin{equation}
\delta E =0, \,  \,\delta p_i = -\omega_{ij} p_j
\end{equation}
Then demanding,
\begin{equation}
\delta \Phi_3 =0
\end{equation}
fixes the transformation of the Grassmann variables under the rotations,
\begin{equation}
\delta \epsilon_i = - \omega_{ij} \epsilon_j, \,\, \delta \eta = \delta \bar\eta =0
\label{gal}
\end{equation}
Correspondingly, the rotation operator (\ref{rotation}) is modified by an additional term. Using (\ref{algebra1}) this is found to be,
\begin{equation}
R = \frac{1}{2} \omega_{ij} (x_i p_j  - x_j p_i - i [\epsilon_i, \epsilon_j])
\label{rotation1}
\end{equation}
which is our cherished expression for the full rotation generator. One may also verify that it is consistent with the relation (\ref{p}) by taking appropriate variations on both sides. The last term, remembering that the $\epsilon_i$ are related to the Pauli matrices (see (\ref{rep})), manifests the spin half nature of the particle.

Proceeding in this manner the complete set of galilean generators in the super phase space are obtained. The boosts, spatial translation and temporal translation are, respectively, generated by,
\begin{eqnarray}
B &=& v_i B_i = v_i( m x_i - t p_i - i\eta \epsilon_i) \cr
P &=& \epsilon_i p_i \cr
T &=& \epsilon E
\label{generators}
\end{eqnarray}

Under these galilean transformations the action ({\ref{g1}) is quasi invariant,
\begin{equation}
\delta S = \int d\tau \,\frac{d}{d\tau}(-m v_i x_i)
\label{quasi}
\end{equation}
Note that for other transformations (rotations etc.) it is exactly invariant. Remember that the same
expression was obtained for the spinless case (see equation ( \ref{quasi1})).

\subsubsection{Gauge Symmetries}

The present model has three first class constraints which signals the occurrence of gauge symmetries.  Following the Dirac conjecture, the generator of these symmetries is  defined as a linear combination of the these constraints,
\begin{equation}
G= \alpha_1\Phi_1 + \alpha_2\Phi_2 + \alpha_3 \Phi_3
\label{generator}
\end{equation}
where the $\alpha$'s are the gauge  parameters. Now it is known that these parameters are not independent. In fact the number of independent parameters is equal to the number of the independent primary first class constraints of the theory \cite{BRR1, BRR2}, which is two here. To find the relation among these parameters we exploit the condition that these parameters must satisfy,
\begin{equation}
\alpha^a \,( g_a\,^{b_2} + \lambda^{a_1} \,f_{a a_1}\,^{b_2}) - \dot\alpha_{b_2} = 0
\label{condition}
\end{equation}
This condition has been discussed in the literature and can be obtained, among other means, by demanding the commutativity of the time differentiation and gauge variation \cite{BRR1}, \cite{BRR2}. Here the label $a$ refers to the complete set of three constraints, the suffix $1\, (2)$ denote the primary (secondary) sector. The structure constants $f$ have been defined in (\ref{constraint algebra}) while the 
other ones $g$ are defined as,
\begin{equation}
\{\Phi_a, H_c\} = g_a\,^b \Phi_b
\label{hamalgebra}
\end{equation}
The only nonvanishing values are given by,
\begin{equation}
g_3\,^1 = -2i\chi , \,\, g_2\,^3 = -1
\label{g}
\end{equation}
Since there is only one secondary constraint, the above equation (\ref{condition})  yields a single condition. Taking $b_2=3$, we find,
\begin{equation}
\alpha^a \,( g_a\,^{3} + \lambda^{a_1} \,f_{a a_1}\,^{3}) - \dot\alpha_{3} = 0
\end{equation}
Putting the appropriate values from (\ref{constraint algebra}) and (\ref{g}), we obtain,
\begin{equation}
\alpha_2 = -\dot \alpha_3
\label{condition2}
\end{equation}
which yields the final form of the generator,
\begin{equation}
G = \alpha_1 \Phi_1 - \dot\alpha_3 \Phi_2 + \alpha_3\Phi_3
\label{generator1}
\end{equation}

A simple way to understand the above structure without going through the general formalism is to realise that such a condition on the gauge parameters is required to establish consistency in the transformation properties of the variables. Let us for instance take the variation on both sides of the equation of motion for $\eta$ given in (\ref{lagrangeequation}),
\begin{equation}
\delta \dot\eta = - 2im \delta\chi
\label{variation1}
\end{equation}
where the variation of $\eta$ and $\chi$ may be separately calculated as,
\begin{eqnarray}
\delta \eta &=& \{\eta, \, G\} = -2im \alpha_3\cr
\delta \chi &=& \{\chi, \, G\} = -\alpha_2
\label{variation2}
\end{eqnarray}
From (\ref{variation1}) and (\ref{variation2}) the condition (\ref{condition2}) follows immediately.

 As already announced there are two independent parameters in the definition of the generator. The parameter $\alpha_1$ is an ordinary variable ( it multiplies the Schroedinger operator) which is linked to the reparametrisation symmetry of the theory, seen earlier. The parameter $\alpha_3$ is Grassmann for similar reasons and is linked to the super gauge transformations. We now discuss them.

The variation of the different variables is found by taking the graded Poisson brackets with the generator (\ref{generator1}). For the coordinates, this is given by,
\begin{equation}
\delta x_i = \{x_i, \,  G\} = \omega \dot x_i + 2\gamma \epsilon_i
\label{x}
\end{equation}
where we have renamed the old parameters $\alpha_1$ and $\alpha_3$ in favour of two new parameters defined as,
\begin{equation}
\omega = \frac{2\alpha_1 m}{\dot t + \chi \eta}, \,\,\,  \gamma = -(\omega \chi - \alpha_3)
\label{redefinition}
\end{equation}
The transformations for all the other variables may be similarly obtained,
\begin{eqnarray}
\delta t &=& \omega \dot t - \gamma \eta\cr
\delta \epsilon_i &=& \omega \dot \epsilon_i +i p_i \gamma\cr
\delta \eta &=& \omega\dot\eta - 2 i m \gamma\cr
\delta \bar\eta &=& \omega \dot{\bar\eta} +i E \gamma\cr
\delta \chi &=& \omega\dot\chi + \dot\gamma +\dot \omega \chi
\label{symmetry}
\end{eqnarray}
where $p_i$ and $E$ are defined in (\ref{momenta}).
In this form it is more transparent to see that the parameters  $\omega$ and $\gamma$, respectively. are connected to the 
 reparametrisation and super gauge symmetries.
 \subsubsection{Action in presence of a magnetic field}
 
 \bigskip

 It is possible to discuss the coupling of this theory to a background magnetic field as an extension of our approach. By following the same arguments as before, the nonrelativistic limit of the Dirac equation in presence of a magnetic field is obtained \cite{BC}. The result is obtainable from 
 (\ref{ps}) by the minimal substitution, $p_i \rightarrow p_i - A_i$, where $A_i$ is the vector potential,
\begin{eqnarray}
E\,\psi - (\sigma.(p- A))\,\eta &=& 0 \cr
(\sigma.(p-A))\, \psi - 2m \eta &=& 0
\label{ps1}
\end{eqnarray}

It can be shown that both components satisfy the following equation,
\begin{equation}
\Big((\sigma.(p-A))^2 - 2m E\Big)\psi = \Big((\sigma.(p-A))^2 - 2m E\Big)\eta = 0
\label{ps2}
\end{equation}
This is basically the Schroednger equation in the presence of a magnetic field. Expansion of the term involving the $\sigma-$matrix immediately leads to the well known $(\sigma.B)$ term, where $B_i$ is the magnetic field. Incidentally the above equation may also be obtained from a minimal prescription of the free Schroedinger equation (\ref{dispersion1}), re-expressed in the form, 
\begin{equation}
(\sigma.p)^2 - 2m E = 0
\label{ps4}
\end{equation}

The appropriate Grassmann operator denoting the Pauli-Schroedinger operator is given by,
\begin{equation}
P= E\eta + 2 (p_i- A_i) \epsilon_i - 2m \bar\eta
\label{LL1}
\end{equation}
and follows from (\ref{LL}) by a minimal substitution. It satisfies,
\begin{equation}
\frac{1}{2}\{ P, P\}_+ = ((\sigma.(p-A))^2
 - 2 m E)
\label{brs21}
\end{equation}  

It is now straightforward to write the first order form for the action,
\begin{equation}
S = \int d\tau \Big(p_i \dot x^i - E \dot t  +\frac{e}{2} (E -\frac{(\sigma.(p - A))^2}{2m}) - \chi ( E\eta + 2 (p_i - A_i) \epsilon_i - 2m \bar\eta) + \frac{i}{2} (\eta \dot{\bar\eta} + \bar\eta \dot\eta) + i\epsilon_i \dot\epsilon_i\Big)
\label{spinaction4}
\end{equation}
In the above construction the Schroedinger constraint is implemented by the multiplier $`e'$ while the Pauli Schroedinger one is done using $\chi$. An advantage of this method is that the constraints are automatically first class as a consequence of the algebra (\ref{brs21}). In other approaches discussed in the literature \cite{G1}, this does not happen due to a lack of a systematic scheme. It is then necessary to modify the constraints to make them first class. Apart from the ambiguities related to the trial and error nature of such a process, the minimal substitution which is inbuilt in our approach is destroyed.
 \section{Action for Non relativistic Particle in Curved Background}
 
  The  parametrized action for a non relativistic  Fermi particle in $3$ dimensional Euclidean space and absolute time has been derived above, in (\ref{spinaction2}).
The action is quasi invariant under the global Galilean transformations (\ref{galtrans}), valid for ordinary variables, augmented by
changes in the spin 'coordinates',
\begin{eqnarray}
\delta \eta &=& 0\nonumber\\
\delta \epsilon^a &=& -{\omega^a}_ b\epsilon ^b +\frac{ v ^ a}{2}  \eta 
\label{galeeo}
\end{eqnarray}
 
 Both (\ref{galtrans})  and (\ref{galeeo}) are obtained from the galilean generators (\ref{rotation1}) and (\ref{generators}).
Note that for spacetime translation the Grassmann variables do not change. This is explicit in the expression of the generators of Galilean transformations (\ref{generators}).

 Now, the invariance of (\ref{spinaction2}) is ensured by the transformations of the fields (\ref{galtrans}), (\ref{galeeo}) and corresponding transformation of their derivatives in a particular way.
Since this is an important point of coupling a theory in flat space with curved 
background,
 we provide below the transformations of the derivatives explicitly,
 \begin{equation}
\delta \dfrac{dx^{0}}{d\tau} = \dfrac{d}{d\tau}(\delta x^{0})=0
\label{der1}
\end{equation}
as $\delta x^0$ is constant,
and 

\begin{equation}
\delta \dfrac{dx^{k}}{d\tau} = -\omega^{k}{}_{j}\dfrac{dx^{j}}{d\tau} - v^{k}\dfrac{dx^{0}}{d\tau}
\label{der2}
\end{equation}
which can be checked easily. The transformations of the derivatives of the Grassman variables can be obtained similarly as,

\begin{eqnarray}
\delta \frac {d{\eta}}{d\tau}& =& 0\nonumber\\
\delta \dfrac{d\epsilon^{k}}{d\tau}& =& -w^{k}{}_{j}\dfrac{d\epsilon^{j}}{d\tau} + \frac{v^k}{2} \dfrac{d \eta}{d\tau} 
\label{globder} 
\end{eqnarray}
The time and space translation parameters along with the rotation and boost parameters are constants, at this stage. We have seen that the corresponding change of the  action (\ref{spinaction2}}) is then a boundary term only,
see (\ref{quasi}). The same equations of motion follow from both the original and the transformed action.
Our theory (\ref{spinaction2}) is, thus, quasi invariant 
 under the global Galilean transformations.

  In the framework of the Galilean gauge theory (GGT) \cite{BMM1}, the coupling with gravity is formulated  in terms of the localisation of the symmetries. The starting point is to introduce local coordinate bases at every point of space time (denoted by $\alpha$), which are trivially connected with the global coordinates (denoted by $\mu$){\footnote{Notation: Indices from the beginning of the alphabet indicate local coordinates while those from the middle, denote global coordinates. Greek symbols correspond to space time while Latin ones correspond to only space.}}
\begin{equation}
x^\mu = \delta^\mu _\alpha x^\alpha
\label{bases}
\end{equation}  
at this stage.  
   But later, we will see that the connection becomes non trivial.
  
The local galilean transformations can be written exactly as (\ref{galtrans}), 
\begin{equation}
x^\mu \rightarrow x^\mu + \zeta^\mu(x, x^0)
\label{localgal}
\end{equation}
and similarly for (\ref{galeeo}) except that the
parameters connected with transformation of the space coordinates are functions of both space and time, whereas $\theta^0$, the time translation parameter is function of time only, keeping in mind the universal nature of time in galilean physics.  Note that the  fields  continue to transform as (\ref{galtrans},\ref{galeeo}), with 
   reference to the local coordinates. But the derivatives cease
   to follow (\ref{der1},\ref{der2}, \ref{globder})
 when the theory is formulated in local coordinates.
   We replace the ordinary derivatives by covariant derivatives where extra gauge fields are introduced. Their transformations under local galilean transformations are assumed so as to ensure that the covariant derivatives transform formally as (\ref{der1}), (\ref{der2})and (\ref{globder}). The theory thus obtained has an immediate geometrical interpretation. Interestingly 
  by localilising the galilean symmetry we obtain the geometrical structure of the NC spacetime \cite{BMM2}. Introduced initially to address some questions regarding the coupling of non relativistic Schrodinger field with background curvature \cite{BMM1}, the method has been applied to a multitude of problems \cite{BMM3, BM4, BM5, BM6, BM7, BMN1,BMN2, BM9} with success. Here, the method will be applied to the spinning particle problem.   

 To localize the symmetry of the action (\ref{spinaction2}) according to GGT ,the ordinary derivatives must be replaced by the covariant derivatives. For the coordinates we have  already constructed the  
appropriate,  covariant derivatives
   $\frac{Dx^{\alpha}}{d\tau}$, corresponding to $\frac{dx^{\alpha}}{d\tau}$ \cite{BMN1}, \cite{BMN2}. Explicitly,
\begin{equation}
\dfrac{Dx^{\alpha}}{d\tau} = \dfrac{dx^{\nu}}{d\tau} \Lambda^{\beta}_{\nu}\partial_{\beta}x^{\alpha} = \dfrac{dx^{\nu}}{d\tau} \Lambda^{\alpha}_{\nu}
\label{red}
\end{equation}
It has been verified \cite{BMN1} that
 the covariant derivatives indeed transform as (\ref{der1}, \ref{der2}) provided  the fields ${\Lambda^\alpha}_\mu$ satisfy the transformations,
 \begin{eqnarray}
 \delta\Lambda^0_0&=&\dot\epsilon\Lambda^0_0\nonumber\\
\delta \Lambda^{a}_{\nu} &=& -\partial_{\nu} \zeta^{\beta}{\Lambda^a}_{\beta} +{ w^a}_{b}{\Lambda^{b}}_{\nu} - v^{a}{ \Lambda^{0}}_{\nu}
\label{lambdatrans1} 
 \end{eqnarray}

 The covariant derivatives of the spin variables will be mooted now.
Since ${\eta}$ is a scalar under rotation and boost, the covariant derivative is same as
the ordinary derivative
\begin{equation}
\dfrac{D\eta}{d\tau} = {{\dot{\eta}}}
\label{covdereta}
\end{equation}
 The only non trivial one is the covariant  derivative of $\epsilon^a$,  to be denoted by
  $\frac{D\epsilon^a}{d\tau}$. We define it as,
\begin{eqnarray}
\dfrac{D\epsilon^{a}}{d\tau} = \frac{d\epsilon^a}{d\tau}+\dfrac{Dx^{\beta}}{d\tau} \Sigma_{\beta}^{\nu}\left(
B_\nu\right)\epsilon^{a}
\label{reducedo}
\end{eqnarray} 
where $\Sigma^\nu{}_\alpha $ is the inverse of ${\Lambda^\alpha}_\mu$,
\begin{equation}
\Sigma^\nu{}_\alpha \Lambda^\alpha{}_\mu = \delta^\nu{}_\mu
\label{inverse}
\end{equation}   

A convenient form of $ \dfrac{D\epsilon^{a}}{d\tau} $ is obtained by using (\ref{red}),

\begin{eqnarray}
\dfrac{D\epsilon^{a}}{d\tau} = \frac{d\epsilon^a}{d\tau}+\dfrac{dx^{\nu}}{d\tau} 
B_\nu \epsilon^{a}
\label{covderepsa}
\end{eqnarray}
The new fields $B_\mu$ are introduced corresponding to the spin degrees of freedom and can be expanded using the generators (\ref{generators}) as,
\begin{eqnarray}
B_\mu = \frac{1}{2}{B_\mu}_{ab}\sigma^{ab} + {B_{\mu a}}\sigma_a
\end{eqnarray}
where, $\sigma_{ab} $
is the spin matrix and $\sigma_a $ follows from the boost generator.  The action of $ B_\mu$ is determined by the gauge principle
\begin{equation}
B_\mu \phi = \delta(\omega)\phi |_{\omega \to - B_\mu \
}
\end{equation}
where, $\omega$ denotes the (infinitesmal) transformation parameter and $\phi$ is in some representation of the gauge group. This immediately gives us 
\begin{equation}
B_\mu x^a =B_\mu{}^a{}_b x^b - B_\mu {}^a x^0
\end{equation}
and
\begin{eqnarray}
B_\mu \eta &=& 0\nonumber\\
B_\mu \epsilon ^a &=& B_\mu{}^a{}_b\epsilon^b - \frac{1}{2}  B_\mu {}^ a \eta
\label{bphi}
\end{eqnarray}


By direct calculations we can show that
the covariant derivatives transform under the local Galilean transformations in the same way as the ordinary derivatives do under global  Galilean transformations , if the
new fields
transform as,
\begin{eqnarray}
 \delta_0\Sigma_0{}^0 &=&  \Sigma_0{}^{\nu}\partial_{\nu}\zeta^0\notag\\ \delta_0\Sigma_0{}^{k} &=& \Sigma_0{}^{\nu}\partial_{\nu}\zeta^{k}+u^a
\Sigma_a{}^k\notag\\ \delta_0\Sigma_a{}^k &=&\Sigma_a{}^{\nu}
\partial_{\nu}\zeta^k + \omega_a{}^b\Sigma_b{}^k \notag\\\delta_0 B_\mu &=&
-\partial_\mu\zeta^{\nu} B_{\nu}+\partial_{\mu}\omega^{ab}\sigma_{ab}
- \partial_{\mu}u^a{\sigma_a}
        \label{lambdatrans} 
\end{eqnarray}
There is a crucial check at this point. 
As a result of their inverse relationship, the transformations given for $ \Lambda^\alpha{}_\nu $ and $
{\Sigma_\alpha{}^\mu }$ in (\ref{lambdatrans1}) and (\ref{lambdatrans})respectively,  must honour  (\ref{inverse}). Direct substitution 
shows that this is indeed the result,
\begin{equation}
\delta\left(\Sigma^\nu{}_\alpha \Lambda^\alpha{}_\mu \right) = 0
\label{inversevar}
\end{equation}
It is nice to see the term by term cancellation in the calculation leading to (\ref{inversevar}).
Remember that these quantities $\delta\Sigma_\alpha ^\mu$ and $\delta \Lambda^\alpha{}_\mu $ have come from
independent algebraic processes of localisation of the Galilean symmetries.
Further, the transformations can be derived from the transformations of the vielbein and its inverse in a curved space time. This is an indication that GGT captures the geometry of NC spacetime.
This issue will be discussed below in details.

The variations of the covariant derivatives in the Grassman sector also satisfy similar forms, as stated above.
 Considering the cardinal significance of this result, it will be instructive to demonstrate the same.
We have already shown the transformations of the ordinary derivatives in flat space and absolute time, in equation (\ref{globder}). According to the GGT algorithm we have to show that  the corresponding covariant derivatives $ \dfrac{D\eta}{{d}\tau} $ and $ \dfrac{D\epsilon^{a}}{{d}\tau} $ will transform  form invariantly i.e
\begin{eqnarray}
\delta \dfrac{D\eta}{{d}\tau}& =& 0\nonumber\\
\delta \dfrac{D\epsilon^{a}}{d\tau}& =& -w^{a}{}_{b}\dfrac{D\epsilon^{b}}{d\tau}+ \frac{1}{2} v^a\frac{D\eta}{d\tau}
\label{required} 
\end{eqnarray}
as dictated by the transformations (\ref{globder}). Considering that the  transformations have already been fixed from the spatial sector one will admit that the claim is non trivial and would like to see the detailed derivation, particularly the second one. Indeed, for the first equation the result follows immediately on substitution of the definition of  $\dfrac{D\eta}{{d}\tau}$ from(\ref{covdereta}) and (\ref{globder}). 
 For the second, we start from the definition (\ref{covderepsa}) and see that the left hand side of (\ref{required}) is given by

\begin{equation}
\delta \dfrac{D\epsilon^{a}}{d\tau}  = \Delta_1{}^a + \Delta_2{}^a
\label{cal}
\end{equation}
where,
\begin{eqnarray}
 \Delta_1{}^a &=& \delta\frac{{d\epsilon}^a}{d\tau}\nonumber\\
 \Delta_2^a &=& \delta \left(\frac{{dx^\mu}}{d\tau} B_\mu \epsilon^a \right)
\end{eqnarray}
To evaluate $ \Delta_2 $ , we have to be careful in calculating the action of $B_\mu$ on $\epsilon $. In a previous section we have seen that $\epsilon^a $ is a vector under rotation. So ,
\begin{equation}
\left[\sigma_{ab}\right]^c{}_d = \left( \delta_{bd}{\delta_a}^ c - \delta_{a d}{\delta_b}^ c\right)
\end{equation}
After some algebra, we get,
\begin{eqnarray}
\Delta_1^a &=&\frac{d}{d\tau} \left(-{\omega^a}_ b\epsilon ^b +\frac{1}{2} v ^ a\eta\right)
\nonumber\\ &=& -\frac {d\omega^a{}_b}{d\tau}\epsilon^b + \frac{1}{2}\frac{d v^a}{d\tau} \eta
-\omega^a_b\frac {d{\epsilon^b}}{d\tau} + \frac{1}{2} v^a\frac{d\eta}{d\tau}\ \label{cal1}
\end{eqnarray}

 The calculation of $\Delta_2^a$ is bit involved. Commuting $\delta $ with $\frac{d}{d\tau}  $   we 
 start, writing the variation as,
 \begin{eqnarray}
\Delta_2^a = \delta \left(  x^{\prime\mu}\right) B_\mu \epsilon^a +  x^{\prime \mu}\delta \left(  B_\mu \epsilon^a  \right)
\label{cal2}
\end{eqnarray}
We will now write from (\ref{cal2}),
\begin{eqnarray}
\Delta_2^a = \zeta^{ \prime \mu} B_\mu \epsilon^a 
&+& x^{\prime\mu}\left(-\partial_\mu \zeta^\nu B_\nu\epsilon^a 
 + \partial_\mu \omega^a{}_b \epsilon^b -\frac{1}{2} \partial_\mu v^a \eta\right)
\nonumber\\ &+&
 x^{\prime\mu} B_\mu \left({- \omega^a}_ b\epsilon ^b +\frac{1}{2}  v^a\eta  \right)
\label{cal22}
\end{eqnarray}
Now $\eta$ is a scalar. So $B_\mu \eta =0$. Using this in (\ref{cal22}) and combining (\ref{cal}), (\ref{cal1}) and (\ref{cal22}) we get the left  hand side of (\ref{required}) as
\begin {eqnarray}
\Delta_1^a + \Delta_2^a
\nonumber\\ &=& -\frac {d\omega^a{}_b}{d\tau}\epsilon^b +\frac{1}{2}\frac{d v^a}{d\tau} \eta
- \omega^a\frac {d{\epsilon^b}_b}{d\tau} +\frac{1}{2} v^a\frac{d\eta}{d\tau}\nonumber\\
&+&
 \zeta^{ \prime \mu} B_\mu \epsilon^a 
+ x^{\prime\mu}\left(-\partial_\mu \zeta^\nu B_\nu\epsilon^a 
 + \partial_\mu \omega^a{}_b \epsilon^b - \frac{1}{2}\partial_\mu v^a \eta\right)
\nonumber\\ &+&
 x^{\prime\mu} B_\mu \left(-{\omega^a}_ b\epsilon ^b +\frac{1}{2}  v^a\eta  \right)
 \end{eqnarray}
 Now, using the chain role,
 \begin{eqnarray}
 \frac{d\Phi\left( x \right)}{d\tau }= \frac{dx^\mu}{d\tau } \partial_\mu \Phi
 \label{chain}
 \end{eqnarray}
Thus we find that
 \begin {eqnarray}
 \Delta_1^a + \Delta_2^a
 &=&-{\omega^a}{}_b\left(\frac{d\epsilon^{ b}}{d\tau} + x^{\prime\mu} B_\mu \epsilon^b\right) 
 +\frac{1}{2} v ^ a \frac{d\eta}{d\tau}
 \end{eqnarray}
Using the definition (\ref{globder}) we find this is just the right hand side of (\ref{required}).

After identifying the proper expressions for the covariant derivatives it is straightforward  to write the locally symmetric theory from (\ref{spinaction2})
as,
\begin{equation}
S = 
\int \left[
\dfrac{1}{2}m \dfrac{ \left(\dfrac{Dx^{a}}{d\tau}- 2\chi\epsilon^a \right)^2}
{\left(\dfrac{Dx^{0}}{d\tau}
+ \chi\eta
\right)} + 2m \chi{\bar{\eta}}+ \frac{i}{2}\left(\eta{\dot{\bar{\eta}}}+ {\bar{\eta}}{\dot{\eta}} \right)+ i\epsilon^a \left( \dot{\epsilon}^a +{ \dot{x}^\beta}{\Sigma_\beta}^\mu B_\mu \epsilon^a  \right)\right] ~d\tau
\label{spinaction22}
\end{equation}

According to GGT this is the action, which is invariant under the general coordinate transformations (\ref{localgal})
in the 
NC background. 
  We explore the geometric connection in the following section.


\subsection{Geometric Connection}

The modified theory (\ref{spinaction22}) is formulated in flat space and time. It is invariant under the local Galilean transformations with respect to local coordinate systems, the connection of which with the global coordinates is at this stage trivial.  We can then form an alternative point of view,where, space time is considered as a four dimensional manifold, charted by the global coordinates $x^\mu $ . The local basis is a non coordinate basis in the tangent space. Following the tenets of galilean gauge theory, the new fields  ${\Sigma_\alpha}^{\nu}$ (${\Lambda_\mu}^{\alpha}$) may be reinterpreted as vielbeins (inverse vielbeins) in a general manifold charted by the  coordinates $x^{\mu}$.
 The local basis  is related with the coordinate basis by the 'vielbeins' ${\Sigma_\alpha}^\mu$
as, 
\begin{equation}
x^\mu = {\Sigma_\alpha}^\mu x^\alpha
\end{equation}
For flat space time, the veilbeins reduce to Kroneckar deltas and we simply reproduce (\ref{bases}). Local symmetries are Galilean whereas the manifold is invariant under the 
diffeomorphism (\ref{localgal}), 
where $\zeta^\mu $ is now interpreted as any well behaved function of $x$ and $t$.
This point of view is buttressed by the fact that the transformations (\ref{lambdatrans})  
carry two set of indices, $\alpha$ designating the local coordinates and $\mu$ designating space time (global) coordinates. Moreover, the transformation equations (\ref{lambdatrans1}, \ref{lambdatrans}) show that
the 
  local symmetry is the Galilean one while the space time transformation is  a diffeomorphism. Observe that the local basis is now related to the coordinate basis by,
  \begin{equation}
  {\hat{e}}^\alpha = { \Lambda^\alpha}^\mu  {\hat{e}}^\mu
  \end{equation}
  So in this reinterpretation, the connection has become non trivial
as we have commented above.
It has been proved elsewhere,  the 4-dim space time obtained in this way above is the Newton-Cartan
manifold \cite{BM4}. 
 The metric is defined as  
 \begin{equation}
h^{\mu\nu}={\Sigma_a}^{\mu}{\Sigma_a}^{\nu}; \hspace{.2cm}\tau_{\mu}={\Lambda_\mu}^{0} =\Theta  \delta_\mu^0
\label{spm}
\end{equation}
and  has the appropriate tensor properties and degenerate form. Likewise,
\begin{equation}
h_{\nu\rho}=\Lambda_{\nu}{}^{a} \Lambda_{\rho}{}^{a}; \hspace{.2cm}\tau^{\mu}=
{\Sigma_0}^{\mu}\hspace{.3cm}
\label{spm2}
\end{equation}
where  ${\Lambda_\mu}^{\alpha}$  is the inverse of  $\Sigma_\alpha{}^\nu $ and $h_{\mu\nu}$ is a second rank covariant tensor. These structures satisfy the standard Newton-Cartan relations
\begin{equation}
h^{\mu\nu}\tau_\nu=0 \,; \, h_{\mu\nu}\tau^\nu=0 \,; \, h^{\alpha \beta} h_{\beta \rho} = {\delta ^\alpha}_\rho - {\tau ^\alpha}\tau_\rho\, ;\, \tau^\mu\tau_\mu=1
\label{b}
\end{equation}
 We will require these 
NC geometric relations in the following analysis.

It is now easy to express (\ref{spinaction22}) in a manifestly covariant form using the Newton Cartan elements. From (\ref{red}), (\ref{covderepsa}) and (\ref{spm2}) we obtain,
\begin{equation}
\dfrac{Dx^{a}}{d\tau}\dfrac{Dx^{a}}{d\tau} = h_{\nu\sigma}\dfrac{dx^{\nu}}{d\tau} \dfrac{dx^{\sigma}}{d\tau} 
\label{red1}
\end{equation}
Also 
\begin{equation}
\dfrac{D\epsilon^{a}}{d\tau} =\left(\dot{\epsilon} ^{ a} +   \dot{x}^{ \mu}B_\mu^{a c}\epsilon^c\right)
\label{red11}
\end{equation}

Using these  the action (\ref{spinaction22}) may be rewritten as,

\begin{equation}
 S 
 = \int \left[\frac{m}{2}\left( \frac{ h_{{\nu\rho}}{\dot x^\nu} \dot x^{\rho} - 4\chi\epsilon^a {\Lambda^a}_\nu \dot x^{\nu}}{\tau_\sigma  {\dot{x}}^\sigma +\chi\eta}\right)
  +2m\chi {\bar{\eta}} + 
  \frac{i}{2}
 \left({\bar{\eta}}{\dot{\eta}} +\eta{\dot{{\bar{\eta}}}} \right) + i\left(\epsilon^a. {\dot{\epsilon}}^a 
+\epsilon^a{\dot{x}}^\mu {B_\mu}^{a c}\epsilon^c 
\right)\right]~d\tau
\label{lagm1}
\end{equation}

   This is the action in manifestly covariant form. Clearly, this  can be interpreted
as the action of a non relativistic  particle coupled with a Newton Cartan background. This coupling is introduced naturally and illustrates the 
strength and efficacy of the algorithm based on Galilean gauge theory.

It is useful to make certain consistency checks on this action. First, let us take the flat limit. The spin connection term $B_\mu$ is dropped. The vierbein $\Lambda$ is replaced by the corresponding Kroneckar delta. The Newton-Cartan structures, defined in (\ref{spm}) and (\ref{spm2}),  are simplified by using $\tau_0=1, \tau_i=0,  h_{0\mu}=0, h_{ij}=\delta_{ij}$. Now one should obtain the action for a spinning particle in flat space. This is indeed so because the action (\ref{spinaction2}) is reproduced.

The next possibility is to take the spinless limit. Now we drop all terms involving Grassmann variables like $\eta, \epsilon, \chi$. Then we find the action,
\begin{equation}
 S 
 = \frac{m}{2} \int\frac{ h_{{\nu\rho}}{\dot x^\nu} \dot x^{\rho}}{\tau_\sigma  {\dot{x}}^\sigma} 
~d\tau
\label{lagm11}
\end{equation}
It reproduces the action for a nonrelativistic free particle coupled to gravity. It was earlier derived by us in {\cite{BMN2}} following the tenets of GGT. As usual, the starting point was the flat space action (\ref{action1}). It is also easy to see this from (\ref{lagm11}) by taking a flat limit. Gauging the global Galilean symmetry of the flat space action led to (\ref{lagm11}).

It is  reassuring to note that the action (\ref{lagm1}) satisfies the relevant consistency checks.

  

\section{Lagrangian Analysis }

 The Euler- Lagrange equation following from (\ref{lagm1})  is,{\footnote{In this section primes denote derivatives with respect to $\tau$.}
 \begin{eqnarray}
 \frac{d}{d\tau}\left(\frac{\partial L}{\partial x'^\mu}\right) - \frac{\partial L}{\partial x^\mu}  =0 
 \end{eqnarray}
 
 We will sketch the calculations in some detail.  The derivatives of $L$ can 
 be straightforwardly computed. Multiplying the overall equation by $h^{\omega\mu} \left( \tau . x^\prime + \chi\eta \right)$ we get,
 \begin{eqnarray}
{ x}^{\prime\prime \omega}&+&\frac{1}{2} h^{\omega \rho}\left(\partial_\nu  h_{\rho\sigma} +
 \partial_\sigma h_{\nu\rho}-  \partial_\rho h_{\nu\sigma}\right)x^{\sigma\prime}x^{\nu\prime}
 \nonumber\\ &-&  \left( \tau^{\omega}\right)  \left ((\tau.{x}^{\prime})^ \prime -
 \tau .{x}^\prime
\frac { \left( \tau . x^\prime +\chi\eta \right){}^\prime} { \left( \tau . x^\prime + \chi\eta \right)} \right)\nonumber\\
&-&\left(x^{\prime\omega}-2\chi\epsilon_a h^{\omega\rho} {\Lambda^a}_\rho \right)
\frac { \left( \tau . x^\prime +\chi\eta \right)^\prime} { \left( \tau . x^\prime + \chi\eta \right)} 
 \nonumber\\&+& 2\chi\epsilon^a h^{\omega\rho}\left(\partial_\rho{\Lambda^a}_\sigma - \partial_\sigma{\Lambda^a}_\rho\right)x^{\prime \sigma} + 2( h^{\omega\rho}\chi\epsilon^a) {\Lambda^a}_\rho )^\prime\nonumber\\
&+& \left(  \frac{1}{2} h^{\omega\rho}h_{\mu\nu}x^{\prime \mu} x^{\prime  \nu }- 2 h^{\omega\rho}
\chi\epsilon^a{\Lambda^a}_\nu  x^{\prime \nu} \right)
   \frac{ \left( \partial_\sigma \tau_\rho - \partial_\rho \tau _\sigma\right)x^{\prime\sigma}}{ \left( \tau . x^\prime + \chi\eta \right)}\nonumber\\
   &+&\left[\left( \frac{1}{m}h^{\omega\rho}\left( B_\rho^{a b}i\epsilon^a\epsilon^b\right)^\prime \right) - {h^{\omega \rho}}\left[\left( \partial_\rho  B_\sigma^{a b} -  \partial_\sigma  B_\rho^{a b}\right)\left(i\epsilon^a\epsilon^b\right)x^{\prime\sigma}\right)\right]\left( \tau . x^\prime + \chi\eta \right)
 = 0
 \label{intermediate}
 \end{eqnarray}
where the abbreviation,
\begin{equation}
\tau.x=\tau_\sigma x^\sigma
\end{equation}
has been used.
 
 We can now introduce the Dautcourt connection \footnote{This means that the spacetime is assumed to be torsionless.},
 \begin{eqnarray}
 \Gamma^\omega{}_{\sigma\beta} = \frac{1}{2} \tau^\omega \left(\partial_\sigma \tau_{ \beta}
+ \partial_\beta \tau_\sigma \right) + \frac{h^{\omega\alpha}}{2}\left(\partial_\sigma h_{\alpha \beta}
 +\partial_\beta h_{\alpha \sigma} -\partial_\alpha h_{\sigma \beta}\right)+ \frac{1}{2} h^{\omega\alpha} \left(K_{\alpha\sigma} \tau_\beta
+ K_{\alpha\beta} \tau_{ \sigma} \right)
\label{d} 
 \end{eqnarray}

 where $K$ is an arbitrary two form.
Now from (\ref{d}) we can write
\begin{eqnarray}
  \frac{h^{\omega\alpha}}{2}\left(\partial_\sigma h_{\alpha \beta}+
 \partial_\beta h_{\alpha \sigma} -\partial_\alpha h_{\sigma \beta}\right) x^{\prime \sigma}  x^{\prime  \beta} &=& \Gamma^\omega{}_{\sigma\beta} x^{\prime \sigma}  x^{\prime  \beta}-\tau^\omega \partial_\sigma \tau_{ \beta}  x^{\prime \sigma}  x^{\prime  \beta}-  h^{\omega\alpha} K_{\alpha\sigma} \tau_\beta  x^{\prime \sigma}  x^{\prime  \beta}
\end{eqnarray} 
We have  the identity,
\begin{eqnarray}
\tau_\alpha^\prime - \partial_\alpha\tau_\sigma x'^\sigma =
\left(\partial_\sigma\tau_\alpha - \partial_\alpha\tau_\sigma\right)x'^\sigma = 0
\label{id}
\end{eqnarray}
because $\left(\partial_\sigma\tau_\alpha - \partial_\alpha\tau_\sigma\right)$ is the temporal component of the torsion tensor and hence is zero
for the torsionless theory. Using this result  in (\ref{intermediate}) and rearranging terms so that the left side takes the usual form of the geodesic, we get the path of a particle falling freely in background gravity as,
 \begin{eqnarray}
 x^{\prime \prime \omega} &+&   \Gamma^\omega{}_{\sigma\beta} x^{\prime \sigma}  x^{\prime  \beta} =
  \frac{{ \left( \tau . x^\prime + \chi\eta\right)}^\prime}{ \left( \tau . x^\prime + \chi\eta \right)}\left(x'^\omega - 2\chi\epsilon^a h^{\omega \rho}{\Lambda^a}_\rho \right)
   -
    \tau^\omega\frac{\left[ \left(\tau \dot  {x}^\prime\right)^\prime \left(\chi\eta  \right) -  \left(\tau \dot  {x}^\prime\right)\left(\chi\eta  \right)^\prime \right]}{\left( \tau . x^\prime + \chi\eta\right)} 
  \nonumber\\   &+&\left[\left( h^{\omega\rho}\left( B_\rho^{a b}\left(i\epsilon^a\epsilon^b\right) \right) ^\prime \left(\tau . x^\prime + \chi\eta \right) - \left(2\Sigma_a{}^\omega  \chi\epsilon^a \right)^\prime \right)
  \right]
  \label{g11}
 \end{eqnarray}
 where  the arbitrary two form automatically gets identified as,
 \begin{eqnarray}
K_{\rho\sigma} &=& \frac{1}{ \tau.x'}\left[2\chi\epsilon^a \left(\partial_\rho{\Lambda^a}_\sigma
- \partial_\sigma{\Lambda^a}_\rho \right)
\right]\nonumber \\ &-& \left[\left( \partial_\rho  B_\sigma^{a b} -  \partial_\sigma  B_\rho^{a b}\right)\left(i\epsilon^a\epsilon^b\right)\left(\tau.x^\prime + \chi\eta \right)
 \right]
\label{K}
 \end{eqnarray}
 Note that in (\ref{g11}), $ \Gamma^\omega{}_{\sigma\beta}$ is completely specified. It is simple to cross check this equation by substituting the Dautcourt connection from (\ref{d}), using the two form (\ref{K}), back in (\ref{g11}). Then (105) is reproduced. By this manipulation we succeed in identifying the two form.

  The above equation successfully gives  results in various limits. It passes to the corresponding equation in the flat limit and agrees exactly with the result for the spinless particle \cite{BMN1} in the limit we put the Grassmann variables
to zero.

\section{Comparison with Existing Results}

This section is devoted to comparing our results with those in earlier literature. At the outset it is useful to recall that these previous 
results were obtained in the context of general relativity and not nonrelativistic curved background that has been discussed here. Nevertheless, as we shall see, it is possible to make a reasonable comparison.

The discussion of a spinning particle in an external gravitational field in the context of general relativity goes back to Mathisson-Papapetrou-Dixon \cite{math, papa, dixon}. In \cite{papa} a multipole expansion was performed around the world line. In the leading (zero order) treatment, only the first moment is retained so that the gradients of the gravitational field are ignored over the complete spatial extension of the body, thereby resulting in a geodesic curve for the coordinates. At the next order of approximation (pole-dipole) the gradients are no longer neglected and there is a deviation from the geodesy.{\footnote{A generalisation of these results till the octupole term was done by Anandan et al \cite{anandan, anandan1}}} The spinning particle problem was also addressed by other authors using various techniques involving Grassmann variables and supersymmetry \cite{ravendal, holten, holten1} as well as finite representations of the Lorentz group \cite{nash} or the introduction of a Routhian \cite{landau} to yield the dynamics of both the coordinates and the spin operator\cite{routhian}.

The results of \cite{papa} are compactly summarised by the three equations \cite{montani},
\begin{eqnarray}
\frac{DP^\mu}{D\tau}&=& \frac{1}{2} R_{\rho\sigma\nu}\,^\mu S^{\rho\sigma}U^\nu\nonumber\\
\frac{DS^{\mu\nu}}{D\tau}&=& P^\mu U^\nu - P^\nu U^\mu\nonumber\\
P^\mu &=& m U^\mu - \frac{DS^{\mu\nu}}{D\tau} U_\nu
\label{papapetrou}
\end{eqnarray}
Here $P^\mu$ and $S^{\mu\nu}$ are the four momenta and the spin operator while $m$ and $U^\mu$ are the mass and four velocity of the particle. The Riemann tensor is denoted by $R_{\rho\sigma\nu}\,^\mu$.
The physical significance of these equations is clear. The first shows a deviation from the geodesic equation due to the presence of an interaction between the spin and the Riemann tensor. The spin tensor evolution shows, from the second equation, a precession around the four velocity. The last relation shows that the four momenta is not aligned with the four velocity. Further, if we contract the last relation by $U_\mu$ and exploit the second relation along with the normalisation $U^\mu U_\mu = c^2$, then we find,
\begin{equation}
P^\mu U_\mu = m c^2
\label{modified}
\end{equation}

 We now compare our results with these equations, discussing how they may be interpreted as the nonrelativistic analogues of (\ref{papapetrou}). Let us first take certain special cases. First, when the spin term is absent but spacetime is curved. Then the geodesic equation is satisfied while the four momenta is aligned with the four velocity. Similar features are revealed for the nonrelativistic case. In our earlier papers \cite{BMN1, BMN2} we have shown that the nonrelativistic free spinless particle in a curved background satisfies the geodesic equation. It was found by adopting the same galilean gauge theory approach used here to discuss the spinning example.
  
 The second possibility is to have a spinning particle in a flat background.  Once again we find from (\ref{papapetrou}) that the geodesic equation must be satisfied, when expressed in terms of the four momenta. The free spinning nonrelativistic particle in a flat background has been discussed here in details in section 2. From the first two equations of (\ref{lagrangeequation}) we see that here also the geodesic equation is satisfied by the momenta, altough the coordnates apparently do not (see (\ref{flatgeodesic})). This happens because the momenta is not aligned with the velocity (see (\ref{momenta})), similar to the lasr relation of (\ref{papapetrou}). Let us next take the second relation of (\ref{papapetrou}). To compare, we first need to define the spin generator. which is given  here by,
 \begin{equation}
 S^{ij}= -i[\epsilon^i, \epsilon^j] = -i(\epsilon^i \epsilon^j - \epsilon^i \epsilon^j)
 \label{spin}
 \end{equation}
that follows from the definition of the angular momentum operator (\ref{rotation1}). It may be recalled that even in the context of general relativity, the spin generator has sometimes  been defined as the antisymmetrised product of two Grassmann variables to discuss the problem of spinning particles \cite{holten1}. Using the equation of motion (\ref{lagrangeequation}) for $\epsilon^i$, we obtain,
\begin{equation}
\frac{dS^{ij}}{dt}= 2 \chi(p^i\epsilon^j - p^j\epsilon^i)
\label{spineq}
\end{equation}
From the definition of the momenta (\ref{momenta}) it is possible to eliminate $\chi\epsilon^i$ in favour of the velocities. We find,
\begin{equation}
\frac{dS^{ij}}{dt}=  (p^i \dot x^j - p^j \dot x^i)
\label{spinequation}
\end{equation}
This is the nonrelativistic analogue of the second relation in (\ref{papapetrou}) for flat spacetime.

The fact that the momenta is not aligned with the velocity, which is the content of the last relation in (\ref{papapetrou}), was already seen here from (\ref{momenta}). We thus finally concentrate on (\ref{modified}). Using (\ref{momenta}) we find,
\begin{equation}
p_i \dot x_i + p_0 \dot x_0 = \frac{m}{2} {\bf v^2}\,;\,v^i=\dot x^i
\label{kinetic}
\end{equation}
where we have exploited the Grassmann property $\chi^2=0$. This is the nonrelativistic analogue of (\ref{modified}).

Let us next consider the final case of a spinning particle in a curved background. The canonical moments $p_\mu$ is obtained from the action given in (\ref{lagm1}),
\begin{equation}
p_\mu =\frac{\partial L}{\partial \dot x^\mu}= m\frac{h_{\mu\rho}\dot x^\rho - 2\chi\epsilon^a\Lambda^a_\mu}{\tau_\sigma \dot x^\sigma +\chi\eta}
-\frac{m}{2}\frac{h_{\nu\rho}\dot x^\nu \dot x^\rho - 4\chi \epsilon^a\Lambda^a_\nu \dot x^\nu}{(\tau_\sigma\dot x^\sigma +\chi\eta)^2}\tau_\mu +i\epsilon^a B_\mu^{ac} \epsilon^c
\label{curvedmomenta}
\end{equation}

 Expectedly we find that it is not aligned with the velocity. To obtain the analogue of (\ref{modified}) we contract it with $\dot x^\mu$ to get,
\begin{equation}
p_\mu \dot x^\mu = \frac{m}{2} \frac{h_{\alpha\beta} \dot x^\alpha\dot x^\beta}{\tau_\sigma \dot x^\sigma} -\frac{1}{2} S^{ij} B_\mu^{ij}\dot x^\mu
\label{curvedrelation3}
\end{equation}
where the spin generator is defined in (\ref{spin}). Note the presence of the coupling between the spin generator and the curvature. The flat space result (\ref{kinetic}) is easily reproduced by setting the curvature term to zero and putting $h_{0\alpha}=0, h_{ij}=\delta_{ij}, \tau_i=0, \tau_0=1$. 

The last term in (\ref{curvedrelation3}) is the interaction that couples the spin with the curvature. It is then possible to define another  momenta by a minimal substitution exactly as happens for a gauge theory where the canonical momenta $p_\mu$  is replaced by the mechanical momenta $P_\mu=p_\mu - A_\mu$. In this case we may as well define the `mechanical' momenta by a minimal substitution,
\begin{equation}
P_\mu= p_\mu + \frac{1}{2}S^{ij} B_\mu^{ij}
\label{newmomenta}
\end{equation}

Then (\ref{curvedrelation3}) takes a very simple form,
\begin{equation}
P_\mu \dot x^\mu = \frac{m}{2} \frac{h_{\alpha\beta} \dot x^\alpha\dot x^\beta}{\tau_\sigma \dot x^\sigma}
\label{curvedrelation} 
\end{equation}
which may be considered as the nonrelativistic analogue of (\ref{modified}).

Finally we come to the structure of the geodesic equation in terms of the momenta. This has to be expressed in terms of the contravariant momenta $p^\mu$ (see (\ref{papapetrou})), which has to be derived from the covariant form $p_\mu$ found earlier. But here we face a problem. The point is that in relativistic physics the indices are raised or lowered using the metric. However, in nonrelaticvistic physics, contrary to the relativistic case, a single nondegenerate metric does not exist. There are two degenerate metrics  discussed and introduced earlier in (\ref{b}). It is possible to exploit both these degenerate structures to define the following contravariant vector from the covariant one,
\begin{equation}
p^\mu = (h^{\mu\nu} + \tau^\mu \tau^\nu) p_\nu = \Omega^{\mu\nu} p_\nu
\label{contravariant}
\end{equation}
The operator $\Omega^{\mu\nu}$ is invertible and is defined as $\Omega_{\mu\nu}$,
\begin{equation}
\Omega_{\mu\nu}=  h_{\mu\nu} + \tau_\mu\tau_\nu\,\,;\,\, \Omega_{\mu\nu}\Omega^{\nu\sigma} = \delta^\sigma_\mu
\label{inverse}
\end{equation}
as may be easily verified upon using (\ref{b}). Likewise the covariant expression may be obtained from the contravariant one by using $\Omega_{\mu\nu}$,
\begin{equation}
p_\mu= \Omega_{\mu\nu} p^\nu
\label{covariant1}
\end{equation}

Employing these relations an attempt can be made to write the equation of motion in terms of $p^\mu$ instead of  (\ref{g11}). Unfortunately, this inversion from $\dot x^\mu$ to $p^\mu$ cannot be carried out completely and we are unable to write the equation of motion in terms of $p^\mu$. Let us mention that a similar situation also arose in a completely different approach \cite{BCG} and the equation of motion was given in terms of the coordinates. We thus refer back to (\ref{g11}) to discuss the geodesic equation. The term proportional to $x^{\prime\omega}$ can be absorbed by appropriately defining the affine parameter. This was done earlier for the spinless case also \cite{BMN2}. The other terms involve the Grassmann variables. There is, however, only one term where a direct interaction between the spin generator and the curvature exists. This is the crucial term that manifests a deviation from the geodesic equation and is also responsible for phenomenological consequences as we discuss at the end of the paper.  A similar trend is also seen in \cite{BCG} where the other terms, besides the interacting one, can be eliminated by a choice of gauge. In our method no gauge fixing has been done.
 \section{Conclusions}
 
 In this paper we have constructed and analysed various aspects of the action for a nonrelativistic (NR) Fermi (spin half) particle moving in either a flat or a curved background.

 The flat theory was obtained by mimicking the steps that led  to the construction of the parametrised form of the action for a spinless theory, which was briefly reviewed.  
For the spinning theory, apart from the Schroedinger mass shell constraint that is the only condition in the spinless case, there is another restriction  given by the Pauli-Schroedinger constraint. This constraint involves Grassmann variables which could also be identified with the spin variables. Implementing both constraints simultaneously, yields a first order form for the action. The second order form for the action follows on eliminating the `momenta'. The galilean transformations for both ordinary and Grassmann variables were obtained from an appropriate generator. The action was shown to be quasi invariant. The two first class constraints of the model were used to construct the generator of the gauge symmetry. It revealed the existence of two independent symmetries- a reparametrisation symmetry and a super gauge symmetry.

The theory in the curved background was obtained by adopting our formalism, known as galilean gauge theory (GGT) \cite{BMM1}, \cite{BMM2}, \cite{BMM3}, \cite{BM4}. The starting point is to consider the theory of the spinning particle in the flat background derived here. The flat space theory is suitably modified so that it has invariance under local galilean transformations. This is done by replacing the ordinary derivatives in the flat space theory by covariant derivatives following a definite principle which is similar to the way covariant derivatives are introduced in order to convert a global gauge symmetry to a local one. The theory so obtained has a geometric explanation since the new fields introduced to convert the ordinary derivatives into covariant ones get identified with the elements of Newton Cartan geometry. Thus we can interpret it as a model for a spinning particle coupled to background Newton Cartan geometry.

Our findings stand good comparison with existing results. Specifically, it was shown in section 5 that, for a spinning particle in a flat background, our results were interpretable as the nonrelativistic analogues of  Papapetrou's relations  given in (\ref{papapetrou}). This was also extended for the curved background, including the violation of the geodesic equation. 
The equation of motion for the cordinates was calculated. It did not follow the geodesic path which agreed with recent literature \cite{BCG}. 
 Since out approach systematically builds up from the flat space action, there are no problems in reproducing the flat space result. This is an important point since there are examples where this limit poses serious problems \cite{SW}. The other feature is that the calculations were directly done at the NR level without taking some limit of the corresponding  relativistic theory in a curved background. Since the limiting prescription is not unique \cite{BGMM},
\cite{BM9} such approaches can suffer from ambiguities. 

These issues also have phenomenological significance in classical and quantum gravity. Propagation of fermions in curved background generates gravitational interaction due to coupling
of its spin with space-time curvature connection. This occurs both in general relativity as well as in the present nonrelativistic scenario.
This interaction leads to an  asymmetry between the left-handed
and right handed partners under CPT transformation. In the case of neutrinos this property can
generate neutrino asymmetry in the Universe. Such a possibility was studied by Mukhopadhyay and Singh \cite{SM} when neutrinos are moving around a rotating (Kerr) black hole. Other possibilities  emanating from the modification of dispersion relation for different helicities of fermions have been considered by Lambiase and Singh \cite{LS}. Such phenemenological consequences could be investigated in the present context since here also we get the coupling term involving the spin generator and the connection.
At a more formal level,  we would like to extend our analysis to include supersymmetry and also strings moving in a nonrelativistic  background.

 

\end{document}
  
 Now we are considering torsion less NC space time.
 So far, we have not used  torsionless condition in our analysis  \footnote {Except in the assumption of the Dautcourt connection which holds for the torsionless NC spacetime}. There are confusions about the complete expression for torsion tensor in non relativistic theories. Analysis using
 GGT
 gives the following formula \cite{BM5},
 \begin{eqnarray}
T^\rho{}_{\mu\nu} &-& h^{\rho \sigma}h_{\nu\lambda}  T^\lambda{}_{\sigma\mu} + h^{\rho \sigma}h_{\mu\lambda} T^\lambda{}_{\sigma\nu}= \tau^\rho\left(\partial_\mu\tau_\nu - \partial_\nu\tau_\mu\right)- h^{\rho\sigma}\left(\partial_\mu h_{\nu\sigma} - \partial_\nu h_{\mu\sigma}\right)
\nonumber\\&+& h^{\rho \sigma}\left( \Lambda_\nu{}^aD_\sigma\Lambda_\mu{}^a -\Lambda_\mu{}^aD_\sigma\Lambda_\nu{}^a
\right) - h^{\rho \sigma}\left[\left(B_\sigma{}^{a0}\Lambda_{\nu a} \Lambda_{\mu 0}-B_\sigma{}^{a0}\Lambda_{\mu a}\Lambda_{\nu 0}\right)\right ]
   \label{condtnn}
\end{eqnarray}
Multiplying both sides of (\ref{condtnn}) by $\tau_\rho$ we get
\begin{equation}
\tau^\rho  T^\rho{}_{\mu\nu}  = \left(\partial_\mu\tau_\nu - \partial_\nu\tau_\mu\right) 
\label{torsion}
\end{equation}
for all values of $\mu$ and $\nu$.If torsion vanishes then we can easily obtain,

\begin{equation}
\left(\partial_\mu\tau_\nu - \partial_\nu\tau_\mu\right) = 0
\label{torsionnew}
\end{equation}

 Note that there is no controversy about the temporal part of the torsion. So (\ref{torsion}) is universally acceptable. Now  consider the spinless particle theory \cite{BMN1},\cite{BMN2}, where
 we have assumed the arbirary part of Dautcourts formula, so as to remove arbitrarines from our description. Applying the torsionless condition, this value of $K$ vanishes.
  There is no reason why it should be otherwise when spin is taken in account \footnote{Note that such choice agrees with similar works \cite{berd} in the literature.}.
We then obtain the condition,
\begin{eqnarray}
 \frac{1}{ \tau.x'}\left[2\chi\epsilon^a \left(\partial_\rho{\Lambda^a}_\sigma
- \partial_\sigma{\Lambda^a}_\rho \right)\left(\partial_\rho {B_\sigma}^{ac}-\partial_\sigma {B_\rho}^{ac}\right)\right] + i\epsilon^a\epsilon^c(\tau.x^\prime + \chi\eta) = 0
\label{Kzero}
 \end{eqnarray}
 One may recall that the equation (\ref{g10}) 
  was shown to pass over to the well known geodesic equation in curved spacetime, when a properly
 scaled affine connection is adopted \cite{BMN2}. It is an interesting issue to see whether the introduction of spin changes the scenario. We can write the equation of motion for $\epsilon^a $
 as,
 \begin{eqnarray}
 \epsilon^{\prime a } = \left[ x^{\prime \mu}B_\mu{}^{ab}\epsilon^b + \frac{m\chi {\Lambda^a}_\nu x^{\prime \nu}}{\left(\tau.x^\prime + \chi\eta \right)} \right]
 \end{eqnarray}
 Multiplying by $\epsilon^c$ from the right we get.

 \begin{eqnarray}
 \epsilon^{\prime a }\epsilon^c = \left[ x^{\prime \mu}B_\mu{}^{ab}\epsilon^b\epsilon^c + \frac{m\chi \epsilon^c{\Lambda^a}_\nu x^{\prime \nu}}{\left(\tau.x^\prime + \chi\eta \right)} \right]
 \end{eqnarray}

 \begin{eqnarray}
 \epsilon^ a \epsilon^{\prime c} = \left[ x^{\prime \mu}B_\mu{}^{cb}\epsilon^a\epsilon^b + \frac{m {\Lambda^c}_\nu \epsilon^a \chi  x^{\prime \nu}}{\left(\tau.x^\prime + \chi\eta \right)} \right]
 \end{eqnarray}
Hence 
\begin{eqnarray}
\left( \epsilon^{ a }\epsilon^c\right)^\prime = \left[ x^{\prime \mu}\left( B_\mu{}^{ab}\epsilon^b\epsilon^c + B_\mu{}^{cb}\epsilon^a\epsilon^b \right) + \frac{m\chi \left( \chi\epsilon^c{\Lambda^a}_\nu + \epsilon^a \chi {\Lambda^a}_\nu\right) x^{\prime \nu} }{\left(\tau.x^\prime + \chi\eta \right)} \right]
 \end{eqnarray}

Using this in (\ref{g11}) and the antisymmetry of $B$ we see that the equation simplifies to

 \begin{eqnarray}
 x^{\prime \prime \omega} +  \Gamma^\omega{}_{\sigma\beta} x^{\prime \sigma}  x^{\prime  \beta} =
  \frac{{ \left( \tau . x^\prime + \chi\eta\right)}^\prime}{ \left( \tau . x^\prime + \chi\eta \right)}\left(x'^\omega - 2\chi\epsilon^a h^{\omega \rho}{\Lambda^a}_\rho \right)
   -
    \tau^\omega\frac{\left[ \left(\tau \dot  {x}^\prime\right)^\prime \left(\chi\eta  \right) -  \left(\tau \dot  {x}^\prime\right)\left(\chi\eta  \right)^\prime \right]}{\left( \tau . x^\prime + \chi\eta\right)} 
 = 0
  \label{g11n}
 \end{eqnarray}
 
 \section{Action for Non relativistic Particle in Curved Background}
 
  The  parametrized action for a non relativistic  Fermi particle in $3$ dimensional Euclidean space and absolute time has been derived above, in (\ref{spinaction2}).
The action is invariant under the global Galilean transformations,
\begin{equation}
x^\mu \rightarrow x^\mu + \xi^\mu
\end{equation}
where,
\begin{eqnarray}
\xi^0 = \delta x^0 &=& -\epsilon\nonumber\\
\xi^k = \delta x^k &=& \epsilon^k -{\omega^k}_j x^j - v^k x^0
\label{galxx}
\end{eqnarray}
following from the definition of the Galilean generators, (\ref{rotation1}) and (\ref{generators}).
These generators also yield the changes in the spin 'co ordinates',
\begin{eqnarray}
\delta \eta &=& 0\nonumber\\
\delta \epsilon^a &=& -{\omega^a}_ b\epsilon ^b +\frac{ v ^ a}{2}  \eta 
\label{galeeo}
\end{eqnarray}
Note that in spacetime translation the Grassman variables do not change. This is explicit in the expression of the generators of Galilean transformations (\ref{generators}).

 Now, the invariance of (\ref{spinaction2}) is ensured by the transformations of the fields (\ref{galxx}, \ref{galeeo}) and corresponding transformation of their derivatives in a particular way.
Since this is an important point of coupling a theory in flat space with curved 
background,
 we provide below the transformations of the derivatives explicitly,
 \begin{equation}
\delta \dfrac{dx^{0}}{d\tau} = \dfrac{d}{d\tau}(\delta x^{0}) = 
-\epsilon^\prime=0
\label{der1}
\end{equation}
as $\epsilon$ is constant.
and 

\begin{equation}
\delta \dfrac{dx^{k}}{d\tau} = -\omega^{k}{}_{j}\dfrac{dx^{j}}{d\tau} - v^{k}\dfrac{dx^{0}}{d\tau}
\label{der2}
\end{equation}
which can be checked easily. The transformations of the derivatives of the Grassman variables can be obtained similarly as,

\begin{eqnarray}
\delta \frac {d{\eta}}{d\tau}& =& 0\nonumber\\
\delta \dfrac{d\epsilon^{k}}{d\tau}& =& -w^{k}{}_{j}\dfrac{d\epsilon^{j}}{d\tau} + \frac{v^k}{2} \dfrac{d \eta}{d\tau} 
\label{globder} 
\end{eqnarray}
The time and space translation parameters along with the rotation and boost parameters are constants, at this stage. We have seen that the corresponding change of the  action (\ref{spinaction2}}) is then a boundary term only,
see (\ref{quasi}). The same equations of motion follow from both the original and the transformed action.
Our theory (\ref{spinaction2}) is, thus, quasi invariant 
 under the global Galilean transformations.

  In the framework of the Galilean gauge theory (GGT) \cite{BMM1}, the coupling with gravity is formulated  in terms of the localisation of the symmetries. The starting point is to introduce local coordinate bases at every point of space time (denoted by $\alpha$), which are trivially connected with the global coordinates (denoted by $\mu$){\footnote{Notation: Indices from the beginning of the alphabet (like $\alpha, \beta$ etc. or a, b  etc.) denote the local basis while those from the middle (like $\mu, \nu$ etc. or i, j  etc.) denote the global basis. Greek indices denote space time while Latin ones denote only space.}}
\begin{equation}
x^\mu = \delta^\mu _\alpha x^\alpha
\label{bases}
\end{equation}  
at this stage.  
  . But later, we will see that the connection becomes non trivial.
  
   Note that the  field  transformations , with 
   reference to the local coordinates,  are formally Galilean  as in flat space i.e. (\ref{galxx}, \ref{galeeo}). But the derivatives cease
   to follow (\ref{der1}, \ref{der2}, \ref{globder})
 when the theory is formulated in local coordinates.
   We replace the ordinary derivatives by covariant derivatives where extra gauge fields are introduced. Their transformations under local galilean transformations are assumed so as to ensure that the covariant derivatives transform formally as (\ref{der1}), (\ref{der2})and (\ref{globder}). The theory thus obtained has an immediate geometrical interpretation. Interestingly 
  by localilising the galilean symmetry we obtain the geometrical structure of the NC spacetime \cite{BMM2}. Introduced initially to address some questions regarding the coupling of non relativistic Schrodinger field with background curvature \cite{BMM1}, the method has been applied to a multitude of problems \cite{BMM3, BM4, BM5, BM6, BM7, BM8, BMN1, BM9, BMN1} with success. Here, the method will be applied to the spinning particle problem. 
  

 To localize the symmetry of the action (\ref{spinaction2}) according to GGT ,the ordinary derivatives must be replaced by the covariant derivatives following the prescription outlined above. Also, care has to be taken about the fact that the transformatin parameters are no longer constants, rather they are functions of space and time.  Keeping in view the universal nature of time in Galilean physics, the time translation parameter is a function of time only while the others are functions of both space and time. As a result we can write
 (\ref{galxx}) simply as,
 \begin{equation}
  x^\alpha \to x^\alpha + \xi^\alpha ,\,\,\, \xi^0 = -\epsilon (x^0), \,\,\,
  \xi^a = \epsilon^a (x, x^0) -\omega^a{}_b (x, x^0) x^b - v^a (x, x^0) x^0 
  \label{localparameter}
  \end{equation} 
 
For the coordinates we have  already constructed the  
appropriate,  covariant derivatives
   $\frac{Dx^{\alpha}}{d\tau}$, corresponding to $\frac{dx^{\alpha}}{d\tau}$ \cite{BMN1}, \cite{BMN2}. Explicitly,
\begin{equation}
\dfrac{Dx^{\alpha}}{d\tau} = \dfrac{dx^{\nu}}{d\tau} \Lambda^{\beta}_{\nu}\partial_{\beta}x^{\alpha} = \dfrac{dx^{\nu}}{d\tau} \Lambda^{\alpha}_{\nu}
\label{red}
\end{equation}
It has been verified \cite{BMN1} that
 the covariant derivatives indeed transform like (\ref{der1}) and (\ref{der2}),
 \begin{equation}
\delta \dfrac{Dx^{0}}{d\tau} = 0
\label{covder1}
\end{equation}
and,
\begin{equation}
\delta \dfrac{Dx^{k}}{d\tau} = -\omega^{k}{}_{j}\dfrac{Dx^{j}}{d\tau} - v^{k}\dfrac{Dx^{0}}{d\tau}
\label{covder2}
\end{equation}
  provided the fields ${\Lambda^\alpha}$ satisfy the transformations,
 \begin{eqnarray}
 \delta\Lambda^0_0&=&\dot\epsilon\Lambda^0_0\nonumber\\
\delta \Lambda^{a}_{\nu} &=& -\partial_{\nu} \xi^{\beta}\Lambda^{a}_{\beta} + w^{a}_{b}\Lambda^{b}_{\nu} - v^{a} \Lambda^{0}_{\nu}
\label{lambdatrans1} 
 \end{eqnarray}

  The consistency of our approach is once again manifest in the fact the same definition  of the  covariant derivatives (\ref{red}) in absence of spin \cite{BMN1} also holds  
   in presence of  spin with identical set  of transformations for the newly introduced fields $\Lambda$.

 The covariant derivatives of the spin variables will be mooted now.
Since ${\eta}$ is a scalar under rotation and boost, the covariant derivative is same as
the ordinary derivative
\begin{equation}
\dfrac{D\eta}{d\tau} = {{\dot{\eta}}}
\label{covdereta}
\end{equation}
 The only non trivial one is the covariant  derivative of $\epsilon^a$,  to be denoted by
  $\frac{D\epsilon^a}{d\tau}$. We define it as,
\begin{eqnarray}
\dfrac{D\epsilon^{a}}{d\tau} = \frac{d\epsilon^a}{d\tau}+\dfrac{Dx^{\beta}}{d\tau} \Sigma_{\beta}^{\nu}\left(
B_\nu\right)\epsilon^{a}
\label{reducedo}
\end{eqnarray} 
where $\Sigma_\alpha{}^\nu $ is the inverse of ${\Lambda^\alpha}_\mu$. 

A convenient form of $ \dfrac{D\epsilon^{a}}{d\tau} $ is obtained by using (\ref{red}),

\begin{eqnarray}
\dfrac{D\epsilon^{a}}{d\tau} = \frac{d\epsilon^a}{d\tau}+\dfrac{dx^{\nu}}{d\tau} 
B_\nu \epsilon^{a}
\label{covderepsa}
\end{eqnarray}
The new fields $B_\mu$ are introduced corresponding to the spin degrees of freedom and can be expanded using the generators (\ref{generators}) as,
\begin{eqnarray}
B_\mu = \frac{1}{2}{B_\mu}_{ab}\sigma^{ab} + {B_{\mu a}}\sigma_a
\end{eqnarray}
where, $\sigma_{ab} $
is the spin matrix and $\sigma_a $ follows from the boost generator.  The action of $ B_\mu$ is determined by the gauge principle
\begin{equation}
B_\mu \phi = \delta(\omega)\phi |_{\omega \to B_\mu \
}
\end{equation}
where, $\omega$ denotes the (infinitesmal) transformation parameter and $\phi$ is in some representation of the gauge group. This immediately gives us 
\begin{equation}
B_\mu x^a =B_\mu{}^a{}_b x^b - B_\mu {}^a x^0
\end{equation}
and
\begin{eqnarray}
B_\mu \eta &=& 0\nonumber\\
B_\mu \epsilon ^a &=& B_\mu{}^a{}_b\epsilon^b - \frac{1}{2}  B_\mu {}^ a \eta
\label{bphi}
\end{eqnarray}


By direct calculations we can show that
the covariant derivatives transform under the local Galilean transformations in the same way as the ordinary derivatives do under global  Galilean transformations , if the
new fields
transform as,
\begin{eqnarray}
 \delta\Sigma_0{}^0 &=&  \Sigma_0{}^{\nu}\partial_{\nu}\xi^0\notag\\ \delta\Sigma_0{}^{k} &=& \Sigma_0{}^{\nu}\partial_{\nu}\xi^{k}+v^a
\Sigma_a{}^k\notag\\ \delta\Sigma_a{}^k &=&\Sigma_a{}^{\nu}
\partial_{\nu}\xi^k + \omega_a{}^b\Sigma_b{}^k \notag\\\delta B_\mu &=&
-\partial_\mu\xi^{\nu} B_{\nu}+\partial_{\mu}\omega^{ab}\sigma_{ab}
+\partial_{\mu}v^a{\sigma_a}
        \label{lambdatrans} 
\end{eqnarray}
There is a crucial check at this point. $ \Sigma_\alpha{}^\mu$ is the inverse of $\Lambda^\alpha{}_\nu$, Hence,
\begin{equation}
\Sigma_\alpha{}^\mu \Lambda^\alpha{}_\mu = \delta^\nu{}_\mu
\label{inverse1}
\end{equation}  
As a result of this inverse1 relationship, the transformations given for $ \Lambda^\alpha{}_\nu $ and $
{\Sigma_\alpha{}^\mu }$ in (\ref{lambdatrans1}) and (\ref{lambdatrans})respectively,  must honour  (\ref{inverse1}). Direct substitution 
shows that this is indeed the result,
\begin{equation}
\delta\left(\Sigma_\alpha{}^\mu \Lambda^\alpha{}_\mu \right) = 0
\label{inversevar}
\end{equation}
It is nice to see the term by term cancellation in the calculation leading to (\ref{inversevar}).
Remember that these quantities $\delta\Sigma_\alpha ^\mu$ and $\delta \Lambda^\alpha{}_\mu $ have come from
independent algebraic processes of localisation of the Galilean symmetries.
Further, the transformations can be derived from the transformations of the vielbein and its inverse in a curved space time. This is an indication that GGT captures the geometry of NC spacetime.
This issue will be discussed below in details.

The variations of the covariant derivatives in the Grassman sector also satisfy similar forms, as stated above.
Considering the cardinal significance of this result, it will be instructive to demonstrate the same.
We have already shown the transformations of the ordinary derivatives in flat space and absolute time, in equation (\ref{globder}). According to the GGT algorithm we have to show that  the corresponding covariant derivatives $ \dfrac{D\eta}{{d}\tau} $ and $ \dfrac{D\epsilon^{a}}{{d}\tau} $ will transform  form invariantly i.e
\begin{eqnarray}
\delta \dfrac{D\eta}{{d}\tau}& =& 0\nonumber\\
\delta \dfrac{D\epsilon^{a}}{d\tau}& =& -w^{a}{}_{b}\dfrac{D\epsilon^{b}}{d\tau}+ \frac{1}{2} v^a\frac{d\eta}{d\tau}
\label{required} 
\end{eqnarray}
subject to the same transformations (\ref{lambdatrans}). Considering that the  transformations have already been fixed from the spatial sector one will admit that the claim is non trivial and would like to see the detailed derivation, particularly the second one. Indeed, for the first equation the result follows immediately on substitution of the definition of  $\dfrac{D\eta}{{d}\tau}$ from(\ref{covdereta}). For the second, we start from the definition (\ref{covderepsa}) and see that the left hand side of (\ref{required}) is given by

\begin{equation}
\delta \dfrac{D\epsilon^{a}}{d\tau}  = \Delta_1{}^a + \Delta_2{}^a
\label{cal}
\end{equation}
where,
\begin{eqnarray}
 \Delta_1{}^a &=& \delta\frac{{d\epsilon}^a}{d\tau}\nonumber\\
 \Delta_2^a &=& \delta \left(\frac{{dx^\mu}}{d\tau} B_\mu \epsilon^a \right)
\end{eqnarray}
To evaluate $ \Delta_2 $ , we have to be careful in calculating the action of $B_\mu$ on $\epsilon $. In a previous section we have seen that $\epsilon^a $ is a vector under rotation. So ,
\begin{equation}
\left[\sigma_{ab}\right]^c{}_d = \left( \delta_{b d}{\delta_a}^ c - \delta_{a d}{\delta_b}^ c\right)
\end{equation}
After some algebra, we get,
\begin{eqnarray}
\Delta_1^a &=&\frac{d}{d\tau} \left(-{\omega^a}_ b\epsilon ^b +\frac{1}{2} 
v^a\eta\right)
\nonumber\\ &=& -\frac{d\omega^a{}_b}{d\tau}\epsilon^b + \frac{1}{2}\frac{d v^a}{d\tau} \eta
-{\omega^a}_b\frac{d\epsilon^b}{d\tau} + \frac{1}{2} v^a\frac{d\eta}{d\tau} \label{cal1}
\end{eqnarray}

 The calculation of $\Delta_2^a$ is bit involved. Commuting $\delta $ with $\frac{d}{d\tau}  $   we 
 start, writing the variation as,
 \begin{eqnarray}
\Delta_2^a = \delta \left(  x^{\prime\mu}\right) B_\mu \epsilon^a +  x^{\prime \mu}\delta \left(  B_\mu \epsilon^a  \right)
\label{cal2}
\end{eqnarray}

  We will now write from (\ref{cal2}),
\begin{eqnarray}
\Delta_2^a = \xi^{ \prime \mu} B_\mu \epsilon^a 
&+& x^{\prime\mu}\left(-\partial_\mu \xi^\nu B_\nu\epsilon^a 
 + \partial_\mu \omega^a{}_b \epsilon^b -\frac{1}{2} \partial_\mu v^a \eta\right)
\nonumber\\ &+&
 x^{\prime\mu} B_\mu \left({- \omega^a}_ b\epsilon ^b +\frac{1}{2}  v^a\eta  \right)
\label{cal22}
\end{eqnarray}
Now $\eta$ is a scalar. So $B_\mu \eta =0$. Using this in (\ref{cal22})and combining (\ref{cal}), (\ref{cal1}) and (\ref{cal22}) we get the right  hand side of (\ref{required}) as
\begin {eqnarray}
\Delta_1^a + \Delta_2^a
\nonumber\\ &=& -\frac {d\omega^a{}_b}{d\tau}\epsilon^b +\frac{1}{2}\frac{d v^a}{d\tau} \eta
- \omega^a\frac {d{\epsilon^b}_b}{d\tau} +\frac{1}{2} v^a\frac{d\eta}{d\tau}\nonumber\\
&+&
 \zeta^{ \prime \mu} B_\mu \epsilon^a 
+ x^{\prime\mu}\left(-\partial_\mu \zeta^\nu B_\nu\epsilon^a 
 + \partial_\mu \omega^a{}_b \epsilon^b - \frac{1}{2}\partial_\mu v^a \eta\right)
\nonumber\\ &-&
 x^{\prime\mu} B_\mu {\omega^a}_ b\epsilon ^b 
 \end{eqnarray}
 Now, using the chain role,
 \begin{eqnarray}
 \frac{d\Phi\left( x \right)}{d\tau }= \frac{dx^\mu}{d\tau } \partial_\mu \Phi
 \label{chain}
 \end{eqnarray}
we easily see that
 \begin {eqnarray}
 \Delta_1^a + \Delta_2^a
 &=&-{\omega^a}{}_b\left(\frac{d\epsilon^{ b}}{d\tau} + x^{\prime\mu} B_\mu \epsilon^b\right) 
 +\frac{1}{2} v ^ a \frac{d\eta}{d\tau}
 \end{eqnarray}
using the definition (\ref{globder}) we find this is just the right hand side of (\ref{required}).

After identifying the proper expressions for the covariant derivatives it is straightforward  to write the locally symmetric theory from (\ref{spinaction2})
as,
\begin{equation}
S = 
\int \left[
\dfrac{1}{2}m \dfrac{ \left(\dfrac{Dx^{a}}{d\tau}- 2\chi\epsilon^a \right)^2}
{\left(\dfrac{Dx^{0}}{d\tau}
+ \chi\eta
\right)} + 2m \chi{\bar{\eta}}+ \frac{i}{2}\left(\eta{\dot{\bar{\eta}}}+ {\bar{\eta}}{\dot{\eta}} \right)+ i\epsilon^a \frac{D\epsilon^a}{d\tau}
 \right]~d\tau
\label{spinaction22}
\end{equation}
where the covariant derivatives have been defined in (\ref{red} and 
(\ref{covderepsa}).
According to GGT this is the action which is invariant under the general coordinate transformations (\ref{localparameter}), 
in the 
NC background. 
  We explore the geometric connection in some detail in the following section.


\subsection{Geometric Connection}

The modified theory (\ref{spinaction22}) is formulated in flat space and time. It is invariant under the local Galilean transformations (\ref{localparameter}) with respect to local coordinate systems. Its connection  with the global coordinates is at this stage trivial. 
 We can now form an alternative point of view,  where space time is considered as a four dimensional manifold, charted by the global coordinates $x^\mu $ . The local basis is a non coordinate basis in the tangent space. Following the tenets of galilean gauge theory, the new fields  ${\Sigma_\alpha}^{\nu}$ (${\Lambda_\mu}^{\alpha}$) may be reinterpreted as vielbeins (inverse vielbeins) in a general manifold charted by the  coordinates $x^{\mu}$. 
 The local basis  is related with the (global) coordinate basis by the 'vielbeins' ${\Sigma_\alpha}^\mu$
as, 
\begin{equation}
x^\mu = {\Sigma_\alpha}^\mu x^\alpha
\end{equation}
Local symmetries are Galilean whereas the manifold is invariant under the diffeomorphism (\ref{localparameter}).
This point of view is buttressed by the fact that the transformations (\ref{lambdatrans})  
carry two set of indices, $\alpha$ designating the local coordinates and $\mu$ designating space time coordinates. Moreover, the transformation equations (\ref{lambdatrans1} , \ref{lambdatrans}) show that
the 
  local symmetry is the Galilean one while the space time transformation is  a diffeomorphism. Observe that the local basis is now related to the coordinate basis by,
  \begin{equation}
  {\hat{e}}^\alpha = { \Lambda^\alpha}^\mu  {\hat{e}}^\mu
  \end{equation}
  So in this reinterpretation, the connection has become non trivial
as we have commented above.
It has been proved elsewhere,  the 4-dim space time obtained in this way above is the Newton-Cartan
manifold \cite{BM4}. 
 The degenerate Newton Cartan structures are defined as  
 \begin{equation}
h^{\mu\nu}={\Sigma_a}^{\mu}{\Sigma_a}^{\nu}; \hspace{.2cm}\tau_{\mu}={\Lambda_\mu}^{0} =\Theta  \delta_\mu^0
\label{spm}
\end{equation}
 and have  the appropriate tensor properties satisfying,
 \begin{equation}
 h^{\mu\nu}\tau_\nu = 0
 \label{NC}
\end{equation}
 Likewise,
\begin{equation}
h_{\nu\rho}=\Lambda_{\nu}{}^{a} \Lambda_{\rho}{}^{a}; \hspace{.2cm}\tau^{\mu}=
{\Sigma_0}^{\mu}\hspace{.3cm}
\label{spm2}
\end{equation}
where  ${\Lambda_\mu}^{\alpha}$  is the inverse of  $\Sigma_\alpha{}^\nu $, satisfy,
\begin{equation}
h^{\alpha \beta} h_{\beta \rho} = {\delta ^\alpha}_\rho - {\tau ^\alpha}\tau_\rho,\,\,\, h_{\mu\nu}\tau^\nu =0
\label{b}
\end{equation}

 We will require these 
NC geometric relations in the following analysis.

It is now easy to express (\ref{spinaction22}) in a manifestly covariant form using the Newton Cartan elements. Using (\ref{red}),  (\ref{covderepsa})and the property  of the vierbeins  (\ref{spm2}) we obtain,
\begin{equation}
\frac{Dx^a}{d\tau}\frac{Dx^a}{d\tau} = h_{\nu\sigma}
\frac{dx^\nu}
{d\tau} \frac{dx^\sigma}{d\tau} 
\label{red1}
\end{equation}

Also 
\begin{equation}
\dfrac{D\epsilon^{a}}{d\tau} =\left(\epsilon ^{\prime a} +   x^{\prime \mu}B_\mu^{a c}\epsilon^c\right)
\label{red11}
\end{equation}
where the prime denotes an ordinary differentiation with respect to the parameter $\tau$.
Using these  the action (\ref{spinaction22}) may be rewritten as,

\begin{equation}
 S =  
  \int  \left[\left(\dfrac{m}{2}\right)\frac{ h_{{\nu\rho}}{x'^\nu} x'^{\rho} - 4\chi\epsilon^a {\Lambda^a}_\nu x^{\prime\nu}}{\tau_\sigma x^{\prime\sigma} +\chi\eta}
  +2m\chi {\bar{\eta}} + 
  \frac{i}{2}
 \left({\dot{\eta}}{\bar{\eta}} +{\dot{{\bar{\eta}}}} \eta\right)+ i\epsilon^a\left( \epsilon^{\prime a} 
+ x^{\prime\mu} {B_\mu}^{a c}\epsilon^c 
\right)\right]~d\tau
\label{lagm1}
\end{equation}

   This is the action in manifestly covariant form. Clearly, this  can be interpreted
as the action of a non relativistic  particle coupled with a Newton Cartan background. Various limis are easily reproduced. For the spinless theory, we simply drop the Grassmann variables so that the action reduces to,
\begin{equation}
 S =  
  \left(\dfrac{m}{2}\right)\int \frac{ h_{{\nu\rho}}{x'^\nu} x'^{\rho} }{\tau_\sigma x^{\prime\sigma}}
  ~d\tau
\label{lagm11}
\end{equation}
which is form given in []. Also, the flat limit of (\ref{lagm1}) trivially reduces to  (\ref{spinaction2}), since the latter was the starting point of our algorithmic process.
 Once again the 
strength and efficacy of  GGT are demonstrated.

  

\section{Lagrangian Analysis }

We start from the  action (\ref{lagm1}) of the non relativistic particle in curved background,
where, as earlier stated, a prime denotes a differentiation with respect to the parameter $\tau$.

 The Euler- Lagrange equation following from (\ref{lagm1})  is,
 \begin{eqnarray}
 \frac{d}{d\tau}\left(\frac{\partial L}{\partial x'^\mu}\right) - \frac{\partial L}{\partial x^\mu}  =0 
 \end{eqnarray}
 
 We will give the calculations in some detail.  The derivatives of $L$ can 
 be straightforwardly computed. Multiplying the overall equation by $h^{\omega\rho} \left( \tau . x^\prime + \chi\eta \right)$ we get,
 \begin{eqnarray}
{ x}^{\prime\prime \omega}&+&\frac{1}{2} h^{\omega \rho}\left(\partial_\nu  h_{\rho\sigma} +
 \partial_\sigma h_{\nu\rho}-  \partial_\rho h_{\nu\sigma}\right)x^{\sigma\prime}x^{\nu\prime}
 \nonumber\\ &-&  \left( \tau^{\omega}\right)  \left(\tau.{x}^{\prime \prime} -
 \tau .{x}^\prime
\frac { \left( \tau . x^\prime +\chi\eta \right){}^\prime} { \left( \tau . x^\prime + \chi\eta \right)} \right)\nonumber\\
&-&\left(x^{\prime\omega}-2\chi\epsilon_a h^{\omega\rho} {\Lambda^a}_\rho \right)
\frac { \left( \tau . x^\prime +\chi\eta \right)^\prime} { \left( \tau . x^\prime + \chi\eta \right)} 
 \nonumber\\&+& 2\chi\epsilon^a h^{\omega\rho}\left(\partial_\rho{\Lambda^a}_\sigma - \partial_\sigma{\Lambda^a}_\rho\right)x^{\prime \sigma} -2 h^{\omega\rho}(\chi\epsilon^a)^\prime {\Lambda^a}_\rho\nonumber\\
&+& \left(  \frac{1}{2} h^{\omega\rho}h_{\mu\nu}x^{\prime \mu} x^{\prime  \nu }- 2 h^{\omega\rho}
\chi\epsilon^a{\Lambda^a}_\nu  x^{\prime \nu} \right)
   \frac{ \left( \partial_\sigma \tau_\rho - \partial_\rho \tau _\sigma\right)x^{\prime\sigma}}{ \left( \tau . x^\prime + \chi\eta \right)}\nonumber\\
   &-&\left[\left( h^{\omega\rho} B_\rho^{a b}\left(\epsilon^a\epsilon^b\right)^\prime \right)+ h^{\omega \rho}\left[\left( \partial_\rho  B_\sigma^{a b} -  \partial_\sigma  B_\rho^{a b}\right)\left(\epsilon^a\epsilon^b\right)x^{\prime\sigma}\right)\right]\left( \tau . x^\prime + \chi\eta \right)
 = 0
 \label{intermediate}
 \end{eqnarray}
where the abbreviation,
\begin{equation}
\tau.x=\tau_\sigma x^\sigma
\end{equation}
has been used.
 
 We can now introduce the Dautcourt connection,
 \begin{eqnarray}
 \Gamma^\omega{}_{\sigma\beta} = \frac{1}{2} \tau^\omega \left(\partial_\sigma \tau_{ \beta}
+ \partial_\beta \tau_\sigma \right) + \frac{h^{\omega\alpha}}{2}\left(\partial_\sigma h_{\alpha \beta}
 +\partial_\beta h_{\alpha \sigma} -\partial_\alpha h_{\sigma \beta}\right)+ \frac{1}{2} h^{\omega\alpha} \left(K_{\alpha\sigma} \tau_\beta
+ K_{\alpha\beta} \tau_{ \sigma} \right)
\label{d} 
 \end{eqnarray}

 where $K$ is an arbitrary two form.
Now from (\ref{d}) we can write
\begin{eqnarray}
  \frac{h^{\omega\alpha}}{2}\left(\partial_\sigma h_{\alpha \beta}+
 \partial_\beta h_{\alpha \sigma} -\partial_\alpha h_{\sigma \beta}\right) x^{\prime \sigma}  x^{\prime  \beta} &=& \Gamma^\omega{}_{\sigma\beta} x^{\prime \sigma}  x^{\prime  \beta}-\tau^\omega \partial_\sigma \tau_{ \beta}  x^{\prime \sigma}  x^{\prime  \beta}-  h^{\omega\alpha} K_{\alpha\sigma} \tau_\beta  x^{\prime \sigma}  x^{\prime  \beta}\nonumber\\ &-&\left( h^{\omega^\rho} B_\rho^{a b}\left(\epsilon^a\epsilon^b\right)^\prime \right)
\end{eqnarray} 
Using this and the identity
\begin{eqnarray}
\tau_\alpha^\prime - \partial_\alpha\tau_\sigma x'^\sigma =
\left(\partial_\sigma\tau_\alpha - \partial_\alpha\tau_\sigma\right)x'^\sigma
\label{id}
\end{eqnarray}
in (\ref{intermediate}) we get the path of a particle falling freely in background gravity as,
 \begin{eqnarray}
 x^{\prime \prime \omega} &+&   \Gamma^\omega{}_{\sigma\beta} x^{\prime \sigma}  x^{\prime  \beta} =
  \frac{{ \left( \tau . x^\prime + \chi\eta\right)}^\prime}{ \left( \tau . x^\prime + \chi\eta \right)}\left(x'^\omega - 2\chi\epsilon^a h^{\omega \rho}{\Lambda^a}_\rho \right)
 - \left(2\chi\epsilon^a\right)^\prime h^{\omega \rho}\Lambda_\rho{}^a   -
    \tau^\omega\frac{\left[ \left(\tau \dot  {x}^\prime\right)^\prime \left(\chi\eta  \right) -  \left(\tau \dot  {x}^\prime\right)\left(\chi\eta  \right)^\prime \right]}{\left( \tau . x^\prime + \chi\eta\right)} 
  \nonumber\\ &-& \left( h^{\omega^\rho} B_\rho^{a b}\left(\epsilon^a\epsilon^b\right)^\prime \right) \left( \tau . x^\prime + \chi\eta \right)
  \label{g11}
 \end{eqnarray}

 where we have identified the arbitrary two form as,
 \begin{eqnarray}
K_{\rho\sigma} &=& \frac{1}{ \tau.x'}\left[2\chi\epsilon^a \left(\partial_\rho{\Lambda^a}_\sigma
- \partial_\sigma{\Lambda^a}_\rho \right)+ i\epsilon^a\epsilon^c(\tau.x^\prime + \chi\eta)\left(\partial_\rho {B_\sigma}^{ac}-\partial_\sigma {B_\rho}^{ac} \right)\right.\nonumber\\&+& \left. \frac{\left(\partial_\rho\tau_\sigma - \partial_\sigma\tau_\rho \right)}{(\tau.x^\prime) + \chi\eta}\left(\frac{1}{2}h_{\rho\beta}x^{\prime\rho}x^{\prime\beta} -2\chi\epsilon^a {\Lambda^a}_\nu x^{\prime \nu}\right) \right]
\label{K}
 \end{eqnarray}
 and used the identity (\ref{id} ).
 .
 Note that in (\ref{g11}), $ \Gamma^\omega{}_{\sigma\beta}$ is completely specified. If we 
drop the spin part the equation of motion for spinless particle is obtained. 
 Compare this equation with th analogous equation of motion for the spinless particle \cite{BMN1},
 \begin{eqnarray}
 x^{\prime \prime \omega} +  \Gamma^\omega{}_{\sigma\beta} x^{\prime \sigma}  x^{\prime  \beta} =
  \frac{{ \left( \tau . x^\prime \right)}^\prime}{  \tau . x^\prime }x^{\prime \omega}
  \label{g10}
 \end{eqnarray}
 and we get exact agreement. Now, equation (\ref{g10}) 
  was shown to pass over to the well known geodesic equation in curved spacetime, when a properly
 scaled affine connection is adopted \cite{BMN2}. The equation to the path (\ref{g11}) can also be written in an alternative form
 
  \begin{eqnarray}
 x^{\prime \prime \omega} &+&   \Gamma^\omega{}_{\sigma\beta} x^{\prime \sigma}  x^{\prime  \beta} =
  \frac{{ \left( \tau . x^\prime + \chi\eta\right)}^\prime}{ \left( \tau . x^\prime + \chi\eta \right)}\left(x'^\omega - h^{\omega \rho}{\Lambda^a}_\rho  2\chi\epsilon^a -\tau^\omega\chi\eta\right) - \tau^\omega\left[\chi\eta\right]^\prime 
  \nonumber\\ &-& \left( h^{\omega^\rho} B_\rho^{a b}\left(\epsilon^a\epsilon^b\right)^\prime \right) \left( \tau . x^\prime + \chi\eta \right)
  \label{g1f}
 \end{eqnarray}
 
  Now we are considering torsion less NC space time.
 So far, we have not used  torsionless condition in our analysis  \footnote {Except in the assumption of the Dautcourt connection which holds for the torsionless NC spacetime}. There are confusions about the complete expression for torsion tensor in non relativistic theories. Analysis using
 GGT
 gives the following formula \cite{BM5},
 \begin{eqnarray}
T^\rho{}_{\mu\nu} &-& h^{\rho \sigma}h_{\nu\lambda}  T^\lambda{}_{\sigma\mu} + h^{\rho \sigma}h_{\mu\lambda} T^\lambda{}_{\sigma\nu}= \tau^\rho\left(\partial_\mu\tau_\nu - \partial_\nu\tau_\mu\right)- h^{\rho\sigma}\left(\partial_\mu h_{\nu\sigma} - \partial_\nu h_{\mu\sigma}\right)
\nonumber\\&+& h^{\rho \sigma}\left( \Lambda_\nu{}^aD_\sigma\Lambda_\mu{}^a -\Lambda_\mu{}^aD_\sigma\Lambda_\nu{}^a
\right) - h^{\rho \sigma}\left[\left(B_\sigma{}^{a0}\Lambda_{\nu a} \Lambda_{\mu 0}-B_\sigma{}^{a0}\Lambda_{\mu a}\Lambda_{\nu 0}\right)\right ]
   \label{condtnn}
\end{eqnarray}
Multiplying both sides of (\ref{condtnn}) by $\tau_\rho$ we get
\begin{equation}
\tau^\rho  T^\rho{}_{\mu\nu}  = \left(\partial_\mu\tau_\nu - \partial_\nu\tau_\mu\right) 
\label{torsion}
\end{equation}
for all values of $\mu$ and $\nu$.If torsion vanishes then we can easily obtain,

\begin{equation}
\left(\partial_\mu\tau_\nu - \partial_\nu\tau_\mu\right) = 0
\label{torsionnew}
\end{equation}

 Note that there is no controversy about the temporal part of the torsion. So (\ref{torsion}) is universally acceptable. Now  consider the spinless particle theory \cite{BMN1},\cite{BMN2}, where
 we have assumed the arbirary part of Dautcourts formula, so as to remove arbitrarines from our description. Applying the torsionless condition, this value of $K$ vanishes.
  There is no reason why it should be otherwise when spin is taken in account \footnote{Note that such choice agrees with similar works \cite{berd} in the literature.}.
We then obtain the condition,
\begin{eqnarray}
 \frac{1}{ \tau.x'}\left[2\chi\epsilon^a \left(\partial_\rho{\Lambda^a}_\sigma
- \partial_\sigma{\Lambda^a}_\rho \right)\left(\partial_\rho {B_\sigma}^{ac}-\partial_\sigma {B_\rho}^{ac}\right)\right] + i\epsilon^a\epsilon^c(\tau.x^\prime + \chi\eta) = 0
\label{Kzero}
 \end{eqnarray}
 
\section{Conclusions}
 
 In this paper we have constructed and analysed various aspects of the action for a nonrelativistic (NR) Fermi (spin half) particle moving in either a flat or a curved background.

 The flat theory was obtained by mimicking the steps that led  to the construction of an action for a spinless theory, which was briefly reviewed.  
For the spinning theory, apart from the Schroedinger mass shell constraint, there is another restriction  given by the Pauli-Schroedinger constraint. This constraint involves Grassmann variables which could also be identified with the spin variables. Implementing both constraints simultaneously, yields a first order form for the action. The second order form for the action follows on eliminating the `momenta'. The galilean transformations for both ordinary and Grassmann variables were obtained from an appropriate generator. The action was shown to be quasi invariant. The two first class constraints of the model were used to construct the generator of the gauge symmetry. It revealed the existence of two independent symmetries- a reparametrisation symmetry and a super gauge symmetry.

The theory in the curved background was obtained by adopting our formalism, known as galilean gauge theory (GGT). Here the starting point is to consider the theory in the flat background found here. The flat space theory is suitably modified so that it has invariance under local galilean transformations. This is done by replacing the ordinary derivatives in the flat space theory by covariant derivatives following a definite principle which is similar to the way gauge fields are introduced in order to convert a global gauge symmetry to a local one. The theory so obtained has a geometric explanation since the new fields introduced to convert the global galilean symmetry to a local one get identified with the elements of Newton Cartan geometry. Thus we can interpret it as a model for a spinning particle coupled to background Newton Cartan geometry.  The equation of motion for the cordinates was calculated. It did not follow the geodesic path which agrees with recent literature. Since out approach systematically builds up from the flat space action, there are no problems in reproducing the flat space result. This is an mportant point since there are examples where this limit poses serious problems. The other feature is that the calculations were directly done at the NR level without taking some limit of the corresponding  relativistic theory in a curved background. Since the limiting prescription is not uinque, such approaches can suffer from ambiguities.  

As possible future work, we would like to extend our analysis to include supersymmetry and also strings moving in a NR background.


\begin{thebibliography}{999}

  \bibitem{BMM1} 
  R.~Banerjee, A.~Mitra and P.~Mukherjee,
  Phys.\ Lett.\ B {\bf 737}, 369 (2014).
  
  \bibitem{BMM3} 
  R.~Banerjee, A.~Mitra and P.~Mukherjee,
  Phys.\ Rev.\ D {\bf 91}, no. 8, 084021 (2015)
  doi:10.1103/PhysRevD.91.084021
  [arXiv:1501.05468 [gr-qc]].
  
   \bibitem{BMM2}R. Banerjee, A. Mitra, P. Mukherjee, Class.\ Quantum Grav.\ {\bf{32}}, 045010 (2015).
 \bibitem{SW}
 D.T. Son and M. Wingate,
Annals.\ of.\ Physics.\ {\bf 321}, {197-224} (2006). 

\bibitem{Grin} 
  B.~Grinstein and S.~Pal,
  Phys.\ Rev.\ D {\bf 97}, no. 12, 125006 (2018)
  
  
 
  

  
  
 \bibitem{Gr} 
  M.~Geracie,
  arXiv:1611.01198 [hep-th].
  
  
    \bibitem{A} 
  A.~Jain,
  Phys.\ Rev.\ D {\bf 93}, no. 6, 065007 (2016)

  \bibitem{Mi} 
  A.~Mitra,
  Int.\ J.\ Mod.\ Phys.\ A {\bf 32}, no. 36, 1750206 (2017)
  
  \bibitem{M} 
  K.~Morand,
  arXiv:1811.12681 [hep-th].
   \bibitem{Rea} 
  J.~Read and N.~J.~Teh,
  Class.\ Quant.\ Grav.\  {\bf 35}, no. 18, 18LT01 (2018) 
  
   \bibitem{K} 
  G.~K.~Karananas,
  doi:10.5075/epfl-thesis-7173 


   \bibitem{Cartan-1923}
E.~Cartan, {\it {Sur les vari\'et\'es \`a connexion affine et la th\'eorie de
  la relativit\'e g\'en\'eralis\'ee. (premi\`ere partie)}},  {Annales
  Sci.Ecole Norm.Sup.} {\bf 40} (1923) 325--412.

\bibitem{Cartan-1924}
E.~Cartan, {\it {Sur les vari\'et\'es \`a connexion affine et la th\'eorie de
  la relativit\'e g\'en\'eralis\'ee. (premi\`ere partie) (Suite).}},  {
  Annales Sci.Ecole Norm.Sup.} {\bf 41} (1924) 1--25.
  
  
    



 

 
  
  
  
  
  
 





   \bibitem{Havas}     
    P. Havas,
Rev. Mod. Phys. {\bf{36}},(1964),938.

\bibitem{Daut}       
    G. Dautcourt:
    {\rm ``Die Newtonske Gravitationstheorie als Strenger Grenzfall
         der Allgemeinen Relativit\"atheorie''},
    Acta Phys. Pol. {\bf{25}},5,(1964),637.



\bibitem{TrautA}     
    A. Trautman, ``Theories of Space, Time and Gravitation" 
    in {\it Lectures on General Relativity},
    S. Deser and K.W. Ford, eds., Prentice-hall, Englewood Cliffs, 1965.
    
  


\bibitem{Kuch}           
   K. Kucha\v{r},
   Phys. Rev. ,{\bf{22D}},6,(1980),1285.




\bibitem{EHL}       
    J. Ehlers:
{\rm ``On Limit Relations between, and Approximative Explanations of,
Physical Theories''},
in {\it Logic, Methodology and Philosophy of Science}, VII,
B. Marcus et al., eds., Elsiever, Amsterdam (1986),405.

\bibitem{PPP}
  R.~De Pietri, L.~Lusanna and M.~Pauri,
  Class.\ Quant.\ Grav.\  {\bf 12}, 219 (1995)
  
  \bibitem{BMN1} 
  R.~Banerjee and P.~Mukherjee,
  Phys.\ Lett.\ B {\bf 797}, 134834 (2019)
  doi:10.1016/j.physletb.2019.134834
  [arXiv:1907.01508 [gr-qc]].

   \bibitem{BGMM}
  R.~Banerjee, S. Gangopadhyay and P.~Mukherjee, Int.J.Mod.Phys. A32 (2017) , 1750115 
  
   \bibitem{j1}
  K,~Jensen,
  SciPost Phys.\  {\bf 5}, no. 1, 011 (2018)

  


  
  \bibitem{ABPR} R. Andringa, E.~A. Bergshoeff, S. Panda and M. de Roo,
  Class.\ Quantum Grav.\  {\bf 28} 105011 (2011).
  
 
  \bibitem{BM10} 
  R.~Banerjee and P.~Mukherjee,
  Nucl.\ Phys.\ B {\bf 938}, 1 (2019)
 
  

  \bibitem{BM4}R. Banerjee, P. Mukherjee, Phys. Rev. {\bf{D93}}, 085020 (2016).

\bibitem{BM5} 
  R.~Banerjee and P.~Mukherjee,
  Class.\ Quant.\ Grav.\  {\bf 33}, no. 22, 225013 (2016)
  doi:10.1088/0264-9381/33/22/225013
  [arXiv:1604.06893 [gr-qc]].   
    
    
\bibitem{BM6} 
  R.~Banerjee and P.~Mukherjee,
  Phys.\ Lett.\ B {\bf 778}, 303 (2018)
  doi:10.1016/j.physletb.2018.01.033
  [arXiv:1710.10882 [gr-qc]].
  
  \bibitem{BM7} 
  R.~Banerjee and P.~Mukherjee,
  Nucl.\ Phys.\ B {\bf 938}, 1 (2019)
  doi:10.1016/j.nuclphysb.2018.11.002
  [arXiv:1801.08373 [gr-qc]].
  
  \bibitem{MS} 
  P.~Mukherjee and A.~Sattar,
  Phys.\ Rev.\ D {\bf 99}, no. 8, 084038 (2019)
  
  
 

\bibitem{BMN2} 

R.~Banerjee and P.~Mukherjee,
Phys. Rev. D \textbf{101}, no.12, 126013 (2020)
doi:10.1103/PhysRevD.101.126013
[arXiv:1910.01452 [gr-qc]].
\bibitem{math} M. Mathisson, Acta Phys. Polon. 6 (1937) 167.

\bibitem{papa} A. Papapetrou, ``Spinning test particles in general relativity I." Proc. Royal Soc. London, A209 (1951) 248

\bibitem{dixon} W.G. Dixon, Proc. Int. School of physics ``Enrico Fermi" LXVII, pp.156 (1979).

\bibitem{anandan} J. Anandan, N. Dadhich and P. Singh, Phys. Rev. D68 (2--3) 124014

\bibitem{anandan1} J. Anandan, N. Dadhich and P. Singh,  Int. Jour. Mod. Phys. D12 (2003) 1651

  \bibitem{HRT} A.~ Hanson, T.~ Regge qnd C.~Teitelboim, {\it ``Constrained Hamiltonian Systems"},
    Academia Nazionale Dei Lincei, Rome, 1976.
    
   \bibitem{BCG}A.~ Barducci,  R. Casselbouni and J,~Gomis, JHEP 1801 (2018) 002.  
   
  \bibitem{LL} J. M. Levy Leblond, Comm. Math. Phys. {\bf{6}}, (1967), 286
  
  \bibitem{BC} R. Banerjee and D. Chatterjee, 
 Nucl.Phys. B954 (2020) 114994
 
 \bibitem{BM9} 
  R.~Banerjee and P.~Mukherjee,
  Phys.\ Rev.\ D {\bf 98}, no. 12, 124021 (2018)
  doi:10.1103/PhysRevD.98.124021
  [arXiv:1810.03902 [gr-qc]].
  
   
   
   \bibitem{ravendal} F. Ravndal, Phys. Rev. D21, 2823 (1980).
   
   \bibitem{holten} R.H. Rietdijk and J.W. van Holten, Class. Quan. Grav. 10, 575 (1992).
   
   \bibitem{holten1}J.W. van Holten, Nucl. Phys. B356, 3 (1991).
   
   \bibitem{nash} P.L. Nash, Jour. Math. Phys. 25, 2194 (1984)
   
   \bibitem{landau} L.D. Landau and E.M. Lifshitz, {\it{Classical Mechanics}}, Pergamon Press, Oxford, 1960, p. 134.
   
   \bibitem{routhian} K. Yee and M. Bander, Phys. Rev. D48, 2797 (1993).
   
   \bibitem{montani} F. Cianfrani and G. Montani, ``Dirac equations in curved spacetime versus Papapetrou spinning particles" arXiv:0810.0447 [gr-qc]
    
  \bibitem{G1} J. Gomis and M. Novell, Phys. Rev. D 33 (1986) 2212 .


  \bibitem{B} E. Bergshoeff, J. Rosseel and T. Zojer, Class. Quant. Grav. 32 (2015) 205003

  
 
  
  
  \bibitem{D}
P. A. M. Dirac, {\it `` Lectures on quantum mechanics " }, Yeshiva University, New york, 1965.
 
   
 \bibitem{BRR1} R.~Banerjee, H.~J.~Rothe and K.~D.~Rothe, Phys. Lett. {\bf{B 479}} (2000) 429- 438, [hep - th/9907217].
\bibitem{BRR2} R.~Banerjee, H.~J.~Rothe and K.~D.~Rothe, Phys. Lett.{\bf{B 462}} (1999) 248- 251, [hep - th/9906072].


\bibitem{SM} B. Mukhopadhyay and  P. Singh, Mod. Phys. Lett. A18 (2003) 779.


\bibitem{LS} G. Lambiase and P. Singh Phys.Lett. B565 (2003) 27-32





\end{thebibliography}

\begin{thebibliography}{999}
 

   \bibitem{Cartan-1923}
E.~Cartan, {\it {Sur les vari\'et\'es \`a connexion affine et la th\'eorie de
  la relativit\'e g\'en\'eralis\'ee. (premi\`ere partie)}},  {Annales
  Sci.Ecole Norm.Sup.} {\bf 40} (1923) 325--412.

\bibitem{Cartan-1924}
E.~Cartan, {\it {Sur les vari\'et\'es \`a connexion affine et la th\'eorie de
  la relativit\'e g\'en\'eralis\'ee. (premi\`ere partie) (Suite).}},  {
  Annales Sci.Ecole Norm.Sup.} {\bf 41} (1924) 1--25.
  
    \bibitem{BM8} 
  R.~Banerjee and P.~Mukherjee,
  Phys. Lett. {\bf B797},
  134484 (2019);
  arXiv:1907.01508 [gr-qc]. 
 



   \bibitem{Havas}     
    P. Havas,
Rev. Mod. Phys. {\bf{36}},(1964),938.

\bibitem{Daut}       
    G. Dautcourt:
    {\rm ``Die Newtonske Gravitationstheorie als Strenger Grenzfall
         der Allgemeinen Relativit\"atheorie''},
    Acta Phys. Pol. {\bf{25}},5,(1964),637.



\bibitem{TrautA}     
    A. Trautman, ``Theories of Space, Time and Gravitation" 
    in {\it Lectures on General Relativity},
    S. Deser and K.W. Ford, eds., Prentice-hall, Englewood Cliffs, 1965.
    
  


\bibitem{Kuch}           
   K. Kucha\v{r},
   Phys. Rev. ,{\bf{22D}},6,(1980),1285.




\bibitem{EHL}       
    J. Ehlers:
{\rm ``On Limit Relations between, and Approximative Explanations of,
Physical Theories''},
in {\it Logic, Methodology and Philosophy of Science}, VII,
B. Marcus et al., eds., Elsiever, Amsterdam (1986),405.

 \bibitem{SW}
 D.T. Son and M. Wingate,
Annals.\ of.\ Physics.\ {\bf 321}, {197-224} (2006). 


   \bibitem{j1}
  K,~Jensen,
  SciPost Phys.\  {\bf 5}, no. 1, 011 (2018)

  

 

\bibitem{PPP}
  R.~De Pietri, L.~Lusanna and M.~Pauri,
  Class.\ Quant.\ Grav.\  {\bf 12}, 219 (1995)
  
  \bibitem{ABPR} R. Andringa, E.~A. Bergshoeff, S. Panda and M. de Roo,
  Class.\ Quantum Grav.\  {\bf 28} 105011 (2011).
  
  
   \bibitem{BMM1} 
  R.~Banerjee, A.~Mitra and P.~Mukherjee,
  Phys.\ Lett.\ B {\bf 737}, 369 (2014).
  
   \bibitem{BMM2}R. Banerjee, A. Mitra, P. Mukherjee, Class.\ Quantum Grav.\ {\bf{32}}, 045010 (2015).
\bibitem{BM5} 
  R.~Banerjee and P.~Mukherjee,
  Class.\ Quant.\ Grav.\  {\bf 33}, no. 22, 225013 (2016)
  doi:10.1088/0264-9381/33/22/225013
  [arXiv:1604.06893 [gr-qc]].   
    
    
\bibitem{BM6} 
  R.~Banerjee and P.~Mukherjee,
  Phys.\ Lett.\ B {\bf 778}, 303 (2018)
  doi:10.1016/j.physletb.2018.01.033
  [arXiv:1710.10882 [gr-qc]].
  
  \bibitem{BM7} 
  R.~Banerjee and P.~Mukherjee,
  Nucl.\ Phys.\ B {\bf 938}, 1 (2019)
  doi:10.1016/j.nuclphysb.2018.11.002
  [arXiv:1801.08373 [gr-qc]].
  
\bibitem{BMM3} 
  R.~Banerjee, A.~Mitra and P.~Mukherjee,
  Phys.\ Rev.\ D {\bf 91}, no. 8, 084021 (2015)
  doi:10.1103/PhysRevD.91.084021
  [arXiv:1501.05468 [gr-qc]].
  
   \bibitem{BM4}R. Banerjee, P. Mukherjee, Phys. Rev. {\bf{D93}}, 085020 (2016).
\bibitem{BMN1} 
  R.~Banerjee and P.~Mukherjee,
  Phys.\ Lett.\ B {\bf 797}, 134834 (2019)
  doi:10.1016/j.physletb.2019.134834
  [arXiv:1907.01508 [gr-qc]].
\bibitem{BMN2} 
  R.~Banerjee and P.~Mukherjee,
  arXiv:1910.01452 [gr-qc].
  \bibitem{kluson} J. Kluson, Eur. Phys. Jour. C78 (2018) 117
  
\bibitem{BM9} 
  R.~Banerjee and P.~Mukherjee,
  Phys.\ Rev.\ D {\bf 98}, no. 12, 124021 (2018)
  doi:10.1103/PhysRevD.98.124021
  [arXiv:1810.03902 [gr-qc]].
  
  \bibitem{B} E. Bergshoeff, J. Rosseel and T. Zojer, Class. Quant. Grav. 32 (2015) 205003

  
  \bibitem{berd}A.~ Barducci,  R. Casselbouni and J,~Gomis, JHEP 1801 (2018) 002.
  
  \bibitem{D}
P. A. M. Dirac, {\it `` Lectures on quantum mechanics " }, Yeshiva University, New york, 1965.
 
   \bibitem{HRT} A.~ Hanson, T.~ Regge qnd C.~Teitelboim, {\it ``Constrained Hamiltonian Systems"},
    Academia Nazionale Dei Lincei, Rome, 1976.
 

 
  
  
  
  
  
Class.\ Quantum Grav.\  {\bf 28} 105011 (2011).
\end{thebibliography}
\end{document}